\documentclass[useAMS,usenatbib]{mn2e}
\usepackage{epsfig}

\usepackage{amsfonts}
\usepackage[fleqn]{amsmath}
\usepackage{amssymb}

\catcode`\@=11
\def\gta{\ifmmode{\,\mathrel{\mathpalette\@versim>\,}}
    \else{$\,\mathrel{\mathpalette\@versim>}\,$}\fi}
\def\lta{\ifmmode{\,\mathrel{\mathpalette\@versim<\,}}
    \else{$\,\mathrel{\mathpalette\@versim<}\,$}\fi}
\def\@versim#1#2{\lower 2.9truept \vbox{\baselineskip 0pt \lineskip
    0.5truept \ialign{$\m@th#1\hfil##\hfil$\crcr#2\crcr\sim\crcr}}}
\catcode`\@=12

\def\Myr{\,{\rm Myr}}

\def\kpc{\,{\rm kpc}}

\def\msun{\,{\rm M}_\odot}

\def\feh{\hbox{[Fe/H]}}

\def\afe{[\alpha/\hbox{Fe}]}

\def\Vc{V_{\rm c}}
\def\Rout{R_{\rm out}}
\def\figref#1{Fig.~\ref{#1}}

\newcommand{\eunit}[1]{\ensuremath{\, \mathrm{#1}}}
\newcommand{\eunitfrac}[2]{\ensuremath{\, \mathrm{#1/#2}}}

\title[Radial Flows in Galactic Chemical Evolution]
{Radial Flows and Angular Momentum Conservation in Galactic Chemical Evolution}

\author[Bilitewski \& Sch\"onrich]{Thomas Bilitewski $^1$\thanks{E-mail:Thomas.Bilitewski@physik.uni-muenchen.de}, Ralph Sch\"onrich$^{1,2}$\\
$^{1}$ Max Planck Institute for Astrophysics, Karl-Schwarzschild-Str. 1, D-85741 Garching \\
$^{2}$ Hubble Fellow, Department of Astronomy, Ohio State University, Columbus, OH 43210 \\
\\
\\
Accepted 2012 July 30. Received 2012 July 5; in original form 2012 May 22 
}

\begin{document}

\pagerange{\pageref{firstpage}--\pageref{lastpage}} \pubyear{2012}

\maketitle

\label{firstpage}

\begin{abstract}
We study the effects of radial flows on Galactic chemical evolution. A simple analytic scheme is developed prescribing the coupling of infall from the intergalactic medium and radial flows within the disc based on angular momentum conservation. We show that model parameters are tightly constrained by the observed \feh-abundance gradient in the Galactic disc. By this comparison the average rotational velocity of the onfalling material can be constrained to $ 0.7 \le v/\Vc \le 0.75$, or respectively $\sim 160 \eunitfrac{km}{s}$ when assuming a constant disc circular velocity of $\Vc = 220 \eunitfrac{km}{s}$. We test the robustness of this value against the influence of other processes. For a very simple model of inside-out formation this value changes only by $\Delta v/\Vc \sim 0.1$, i.e. $\sim 20 \eunitfrac{km}{s}$, and significantly less on more realistic scenarios, showing that inside-out formation does not alone explain the abundance gradient. Effects of other uncertain parameters, e.g. star formation 
history and star formation efficiency have very small impact. 

Other drivers of inflow beyond our explicit modelling are assessed by adding a fixed inflow across the whole disc. The churning amplitude only mildly affects the results mostly by slightly flattening the metallicity gradient in the inner disc. A new process causing radial gas flows due to the ejection of material by stars moving on non-circular orbits is studied and seems to contribute negligibly to the total flows. 

We further show that gaseous outer discs cannot be the main source feeding the persistent star formation in the inner regions by a direct inflow.
\end{abstract}


\section{Introduction}
Radial flows of gas have been studied extensively in models of chemical evolution. Their importance has first been recognised by \cite{TinsleyLarson1978}, who emphasise their effects on galactic abundance gradients and conclude that these cannot be ignored in any viable model. \cite{LaceyFall1985} discussed three main drivers of these flows.
First, onfalling gas might have a different specific angular momentum than the gas rotating in the disc. As the onfalling gas is suspected to have a lower rotational velocity, these flows are expected to be directed inwards at a velocity in the range of a few\,\eunitfrac{km}{s} \citep{LaceyFall1985}.
Second, friction within the rotating gas layers might cause radial flows. A viscous disc is subject to an angular momentum flux whenever the angular speed changes with radius. A classic Galactic disc with its nearly flat rotation curve has a global outwards directed angular momentum flux, which drives inflows in the inner regions and an expansion of its outskirts. Estimates of this effect yield velocities of the order of 0.1\eunitfrac{km}{s} \citep{LaceyFall1985}.
Third, interactions of the gas with a central bar and spiral patterns, especially around the corotation resonance, will cause radial flows, typically directed away from the position of the resonance. \citet{LaceyFall1985} cite velocities of 0.3\eunitfrac{km}{s} at a distance of 10\eunit{kpc} from the galactic centre. Further the gas loses angular momentum to the stellar population by interactions/shocks along the spiral arms. A comparison of these quantities shows that especially stronger flows are preferentially driven by angular momentum dilution via onfall.

Infall has been known to affect the evolution of disc galaxies for many years \citep{Larson1972}. Even if no gas at all was lost to the intergalactic medium, roughly one third of the mass involved in each star formation cycle would get locked up in stellar remnants or very long-lived low mass stars and the star formation in spiral galaxies would exhaust the available gas reservoirs on cosmologically short timescales \citep{Larson1980}. Hence the observed star formation rates require ample infall of fresh gas. Infall is also required from a chemical evolution point of view to avoid exceedingly high metallicities after a few Gyrs and to avoid deuterium abundances far below the observations \citep{Linsky2006}. The mode of this accretion is obscure: It is frequently adopted that gas is delivered directly to galaxies by infalling satellites. This direct infall is observed \citep{Sancisi2008}, but they also find that it brings about an order of magnitude too little material to 
sustain the current star
formation rates. Further, such ``cosmological infall'' leads to
rather unsatisfacory outcomes \citep[see][]{Colavitti08}
 in chemical evolution models, since the clumpy accretion and star formation produce unobserved excursions in the detailed abundances, especially in the $\feh-\afe$ plane.

Analytical chemical evolution models assumed a vanishing angular velocity of onfalling material \citep{MayorVigroux1981}. Later onfalling material with a constant fraction of the disc's angular velocity at the given radius was considered as well \citep{PittsTayler1989,PittsTayler1996}. But no argument aside from trying to fit the observed abundance gradients could be made for any particular prescription. In a different line of development radial gas flows were implemented to classical chemical evolution models, starting with \cite{GoetzKoeppen}, later \cite{Portinari00} and recently by \cite{Spitoni11}. Those models assume preset patterns of the inflow velocity in the disc, but do not draw a significant link to physical properties like the angular momentum budget.

It seems natural, though, to link the available angular momentum budget to the puzzle of Galactic accretion. The deficit of visible cold gas infall was recently connected to the presence of a concurrent mode of delivery: Hot coronal gas can be swept up by turbulent cooler gas clouds that are driven out of the Galactic plane by supernova feedback \citep{FB06,FB08,MB10}. Dynamic arguments and observations strongly suggest that the coronal gas should lag behind the rotation speed of the disc \citep{Marinacci2011}. Theoretical calculations for the Milky Way predict values for the global accretion rate sufficient to sustain its star formation and put an estimate of the lag of accreted coronal material at around $75 \eunitfrac{km}{s}$ compared to the disc rotational velocity of $220 \eunitfrac{km}{s}
$ \citep{Marasco2011}. Hence in contrast to cosmological cloud accretion this offers a firm framework with relatively well constrained
properties of accretion. In addition they derive a specific profile for the accretion of gas onto the Galactic disc.

A process neglected so far is the systematic lag of stellar yields themselves behind the local gas velocities as well as friction between the stars and the interstellar medium. This is mediated by the larger velocity dispersion and hence asymmetric drift of stars. We will discuss this in the following paper.

The purpose of this paper is to give a better parametrisation of flows than that used originally in the model of SB09 in the sense that it provides sufficient freedom to fit the data and depends on physically meaningful parameters. The model crucially relies on the assumption of conservation of angular momentum which may not be true in all cases, but which should be a plausible first approximation to understand the overall flows in the Galaxy. Our approach is different from those found in the literature as we assume a radially varying profile for the angular velocity of the onfalling gas. Fitting the model to the data provides information on the kinematic properties of the onfalling gas. These properties depend on the proposed origin of the onfalling material and the assumed way of accretion onto the Galaxy. A comparison of the predictions derived from a model of chemical evolution with those of theories for the origin of the onfalling material might help to shed light on the question of how galaxies accrete
gas.

The paper is organised as follows. Section \ref{sec:SB09} summarises the aspects of the SB09 model \citep{SBI} important to this work. In section \ref{sec:modeleq} we give the equations of the new flow model to be used in the simulations. We model the angular velocity of the onfalling gas as a linear function of galactocentric radius. The general behaviour of the model and the dependence on its parameters is presented in section \ref{sec:modelbe}. In section \ref{sec:Results} we fit the new model to the Galactic radial abundance gradient to constrain the physical model parameters. We investigate the implications of the accretion profile proposed by \cite{Marasco2011} in section \ref{sec:accprofile}. In section \ref{sec:dyingstars} we study the proposed new mechanism for causing radial flows due to the lag of stellar yields. Section \ref{sec:inside-out} discusses an implementation of inside-out formation in the model and section \ref{sec:origin} explores limits on the accretion from an extended gas disc. 
Section \ref{sec:sum} sums up.

\section{The underlying model}\label{sec:SB09}
As a starting point we use the model of \cite{SBI} (SB09). In the following section we briefly recapitulate those parts of the model that are of particular importance to this work.

\subsection{The setup}
The simulation follows the evolution of a disc of size 20\eunit{kpc} divided into 80 annuli, which are equispaced with a width of 0.25\eunit{kpc}. The model has discrete timesteps of 30\eunit{Myrs}. \\
Each annulus contains a cold gas phase, which produces stars, and a hot phase, which is fed by stellar yields and condenses back onto the cold phase. The model tracks the detailed abundances in each element considered and holds information over the stellar populations formed at each timestep.

\subsection{Star formation rate and initial mass function}
The star formation rate (SFR) is assumed to follow a simple Schmidt law \citep[][]{Schmidt59}, i.e. it depends on the surface density of the gas (${\Sigma}_{g}$) according to a power law, whose index was determined by \citet{Kennicutt1998}. We hence use
\begin{equation*}
\frac{d{\Sigma}_{star}}{dt}= k \begin{cases}{\Sigma}^{1.4}_{g} &{\Sigma}^{1.4}_{g} > {\Sigma}_{c} \\
					     C\,{\Sigma}^{4}_{g} & \text{else}  $,$
\end{cases}
\end{equation*}
where C is chosen such that the SFR is continuous in $\Sigma$ and $k$ is fitted to observations. For simplicity we assume that the critical density ${\Sigma}_{c} = 0$ throughout this work. A positive value would produce a star formation break in the Milky Way, beyond which the star formation rate drops sharply. Such a break should not take place out to $12 \kpc$ and beyond this point it has only very minor effects on the local chemistry. \\
For the initial mass function (IMF) the model uses a Salpeter function with upper and lower mass limits of $0.1 \, \msun$ and $100 \, \msun$
\begin{equation*}
  \frac{dN}{dM} \propto M^{-2.35} \, .
\end{equation*}
Apart from some influence on detailed abundances (which are of no interest for this work) by the shape of the IMF at high solar masses, it mainly impacts the lock-up of material at lower masses and by this the need for fresh material to sustain a certain star formation rate and influences the general metallicity level in the disc. In good agreement with general findings we assume that there is at best a weak dependence of the IMF on Galactocentric radius.

\subsection{Flows and onfall}
The disc is initialised in the first time-step with a gas mass of $M_0$ and an exponential surface density profile. Subsequent onfall increases the gas mass of the disc by
\begin{equation*}
 \dot{M}=\frac{M_1}{b_1}\, e^{-t/b_1} + \frac{M_2}{b_2} e^{-t/b_2} \, $.$
\end{equation*}
Starting from a relatively low initial mass $M_0$, the double exponential law allows for a rapid build-up of the Galactic disc mass and SFR on the short timescale $b_1$, followed by a slower decline set by the long timescale $b_2$. Such an onfall law with two timescales has been widely used in Galactic models starting with \cite{Chiosi1980}. In contrast to models without radial flows, in which the radial structure of onfall can be inferred from the present mass densities, only the total accreted mass can be constrained in our model at this point.

In general the mass balance at each ring then reads
\begin{equation}
\dot{M}_i = (O_i - D_i) + (T_i - \Psi_i)   + \Delta_i,
\end{equation}
with onflow $O$, outflow/loss to the IGM or to the corona $D$, star formation $\Psi$, stellar yields $T$ and the redistribution balance $\Delta$ from mass exchange with neighbouring annuli.

To preserve an exponential disc profile we hence have two free functions to fulfill a single constraint: The spatial distribution of infall and the redistribution of material throughout the disc. To deal with this underdetermination SB09 used a very simple parametrisation consisting of the two parameters $f_a$ and $f_b$, which determine the shape of radial inflow and infall from the IGM. As a main ingredient of this paper we will replace this simple parametrisation with the far more physical condition of angular momentum conservation and thus trace the redistribution of material back to angular momentum exchange and the specific angular momentum of onfall.\\

\subsection{Kinematics}
To cope with Galactic dynamics and kinematics, the model includes a machinery to calculate the spatial distribution and kinematic of stars using a radius-dependent age-dispersion relationship under the assumption of isothermal subpopulations. More importantly it parametrised radial migration in gas and stars via an effective exchange of mass between different annuli in concordance with the mechanism of \cite{SellwoodB01} and by reproduction of the local metallicity distribution proved the existence of significant radial migration in our Galaxy.
The mass exchange happens with adjacent rings at each sub-timestep, allowing for two exchanges per time-step. The parametrisation depends on the churning amplitude, a free parameter of the model, which in this work we keep at the preferred value of $k_{ch}=0.25$ if not explicitly stated otherwise. In any case slightly stronger or weaker mixing of the gas and stellar yields has a minor impact on the metallicity gradient of the gas.
The value of SBI was derived on the assumption a radial abundance gradient that matched a blend of different datasets throughout the Galactic disc. This gradient dominates the width of the local abundance distribution, so that a shallower gradient needs to be balanced by higher churning rates.

\section{New flow model equations}\label{sec:modeleq}
In this section we outline how we drive radial flows through the Galactic disc by angular momentum conservation with respect to the onfall of fresh material from the intergalactic medium (IGM) or the Galactic corona.

In the following we will assume a flat rotation curve, i.e. a fixed circular speed $\Vc$. While this is anyway a good approximation \citep{Foster2010}, moderate deviations will have very weak impact on our results, espcially since we will define the velocity of infalling material in units of the local circular speed. We further assume in concordance with the observations that the gas is on nearly circular orbits.

Three functions in Galactocentric radius $R$ are of importance to the infall problem: The surface density of gas already present in the Galactic disc $\Sigma_p(R)$, the specific angular momentum or respectively azimuthal velocity of the onflowing/accreted material $v_o(R)$ and the surface density flux of the onfalling material $s_o(R)$. $\Sigma_p(R)$ is essentially known from the model history, while there are hardly any constraints on $v_{o}(R)$ and $s_o(R)$. Only the total amount of infall is preset by the model and hence the problem is underconstrained by the only constraint of the surface density profile after infall. As in SBI we can now use measured abundance gradients to constrain the functions. To keep the number of free parameters as low as possible while allowing for some freedom we assume a linear dependence of $v_o(R)$ on the Galactocenric radius and parametrise
\begin{equation}\label{eq:keyparam}
v_{o}(R) = \Vc\left(b + a\frac{R}{R_{out}}\right) \, ,
\end{equation}
where we assume an outer disc radius $R_{out} = 20 \kpc$ fit to the limit of the calculation, $\Vc$ is the circular speed of the Galaxy, $b$ is the central ($R = 0$) relative angular momentum of the accreted material and $a + b$ is the relative angular momentum at the outer edge. Once $v_{o}(R)$ is known, the onfall density profile $s_o(R)$ is fully determined by the aimed at gas surface density.

In the following we will typically consider the case that the angular momentum of onfalling gas is lower than that of the circular orbit and speak of radial inflow even if radial outflow would occur if the angular momentum of the onfalling material were higher than that of the disc gas. This is not strictly necessary for the general formalism introduced here, but covers the parameters relevant to this work and  allows a simple scheme of solving the equations iteratively.

For convenience we integrate the onfall over a small time (our timestep) to obtain $\Sigma_o(R)$ the surface density of material dumped over that time onto the disc. In each annulus this dilutes the angular momentum causing a deficit in relative angular momentum $\Delta J_{o}$ compared to a circular orbit:
\begin{equation}
 {\Delta J}_{o}=2\pi \int_{R_{n}}^{R_{n+1}}{ \left( v_{o}(R)-\Vc(R)\right)R \,  \Sigma_{o}(R,t) R\,dR} \, ,
\end{equation}
where $v_{d}(R)$ is the rotation curve of the Galaxy and $\Sigma_{o}(R,t)$ is the area density of the onfalling gas. Assuming that the surface density of onfalling material is a smooth function and considering the relatively fine spacing of the model, we can well approximate the onfalling mass onto each ring as $M_{o, n} \sim \pi \Sigma_{o}(R_{n+1}^2 + R_n^2)$ by neglecting the spatial variation of $\Sigma_o$ over the bin. Moving this out of the integral, we hence obtain:
\begin{align}
 {\Delta J}_{o} = \frac{2\,M_{o, n}(\bar{R},t)}{R_{n+1}^{2}-R_{n}^{2}} \int_{R_{n}}^{R_{n+1}}{ \left( v_{o}(R)-v_{d}(R)\right) R^{2} dR} \, .
\end{align}
With the assumed flat rotation curve $v_d(R)=\Vc$ and using the parametrisation of equation (\ref{eq:keyparam}) the integral solves to:
\begin{multline}
  \int_{R_{n}}^{R_{n+1}}{ \left( v_{o}(R)-\Vc\right) R^{2} dR}= \\
 = \Vc  \left( \frac{a}{4\Rout} ( R_{n+1}^{4}- R_{n}^{4}) \right.+
 \frac{b-1}{3} (R_{n+1}^{3} -R_{n}^{3}) \Big)
\end{multline}
Setting an equidistant spacing with width $\Delta R$ for the $N$ rings, we get
\begin{multline}\label{eq:onfallmomentum}
{{\Delta J}_{o}} = \left(\frac{2a}{4N} \frac{4n^{3}+6n^{2}+4n+1}{2n+1}\right. + \\
\left. + \, \frac{2(b-1)}{3}\frac{3n^{2}+3n+1}{2n+1}\right) \Delta R \Vc \, M_{o}(R,t) \, .
\end{multline}
Now we can make up the angular momentum balance compensating the angular momentum loss caused by onfall via an inflow through the disc
\begin{equation}\label{eq:Angularconservation}
0= {\Delta J}_{tot}(R) = {\Delta J}_{o}(R) + {\Delta J}_{i}(R) \, ,
\end{equation}
where
\begin{equation}
{\Delta J}_{i}(R)= M_{i}(R,t) \Vc \Delta R
\end{equation}
is the angular momentum given up by the radial flow of disc gas and $M_{i}(R,t)$ is the mass flowing in from the next ring.
Plugging equation~(\ref{eq:onfallmomentum}) into equation~(\ref{eq:Angularconservation}) yields
\begin{multline}
   \frac{M_{i}(R,t)}{M_{o}(R,t)}=\left(\frac{-2a}{4N} \frac{4n^{3}+6n^{2}+4n+1}{2n+1} \right.+ \\
                         \left.\frac{2(1-b)}{3}\frac{3n^{2}+3n+1}{2n+1}\right) \label{eq:Massquotient}
\end{multline}
One might be concerned about the explicit factors of $\Delta R$ and $N$ possibly making the results of the simulation resolution dependent. However, for our choice of the angular momentum of onfalling gas equation~(\ref{eq:Massquotient}) is equivalent to expression 7 in the paper by \cite{LaceyFall1985} as can be easily seen by using the relation between masses ($M_{o/i}$, respectively for onfalling and inflowing gas), their surface densities ($\sigma_o$, $\sigma_g$ for onfalling and disc gas) and the inflow velocity ($v_i$).
\begin{align}
M_{i}(R,t)   &= 2 \pi \sigma_{g} R v_i \Delta t \\
M_{o}(R,t) &= 2 \pi \Sigma_{o}  R \Delta R \\
v_{i}(R,t) &= \frac{M_i}{2\pi \sigma_{g} R \Delta t} =- \frac{\dot{\sigma}_{o}}{\sigma_{g}} \frac{R(v_{o}-v_{d})}{v_{d}} \label{eq:inflowspeed}
\end{align}
Holding a discretised version of these expressions, we test that our result is not affected significantly by resolution problems in the appendix, section \ref{subsec:reseff}.

Equation~(\ref{eq:Massquotient}) couples flows and infall by the conservation of angular momentum. This coupling combined with the total need for fresh gas in each ring determines the total amount of mass falling onto each ring from the IGM and radially flowing from the next outer ring. Here we consider exclusively the case of lower angular momentum of the onfalling gas, i.e. the gas will always flow radially inwards. As in this scheme each ring just receives mass from its outer neighbour, it can be solved in a simple way running against the flow direction, i.e. iteratively from the centre to the outermost ring (like the scheme of SBI). This procedure is only viable as long as the flow has a single direction throughout the whole Galaxy, otherwise a more sophisticated method would be required. The outermost ring is assumed to be supplied by a gas layer extending beyond that of the stellar disc modelled with 
the metallicity of the IGM.

The presented scheme fails if the mass of the next outer ring is not sufficient to cover the required inflow. In such a case we limit the inflow to the total mass of this ring, which is equivalent to increasing the momentum of the infalling gas. This procedure is somewhat arbitrary, and the problem could be straight-forwardly tackled by finer timesteps. Yet, it sets a natural limit to the maximum flow velocities of $\sim 8.2\eunitfrac{km}{s}$ within the constraints of 5-10\eunitfrac{km}{s} from observations of external galaxies \citep{Wong2004}. Besides being a plausible limit, at our examined parameter space it bears relevance only at the earliest timesteps of the simulation when the disc builds up rapidly. For the inflow model to make sense, this adjustment should not be necessary for later times in the simulation. We made sure that our choice does not affect the results as compared to a finer time resolution.

In the common case onfalling gas should in all annuli have less or equal angular momentum than the disc gas, which constrains $a + b \le 1$.
Of special interest are the case of constant azimuthal velocity $a=0$, which has been studied in the literature, and the case of rigid rotation with $b=0$. We also point out that a fixed relative angular momentum of onfalling material is by no means synonymous to a constant inflow speed in the disc: Even if the onfalling mass has the same radial density distribution the same inflow speed needs a higher angular momentum difference as evident from equation~(\ref{eq:inflowspeed}).

Via the angular momentum conservation the flows are more tightly constrained than in the original SB09 model. They now depend on physical properties that are principally observable and connected to physical processes. But even without any observational constraints, the chosen parametrisation leaves less freedom, for example it is not possible to provide the total gas mass by radial flows alone, whereas pure infall is still an option.

This treatment is extremely simplified and does in no way describe the whole complexity of the processes involved. It would be naive too believe that the accreted material has some smooth and fixed angular momentum. We should hence bear in mind that the laid out formalism rather gives effective profiles resulting from all processes that may affect the gas before it mixes with the disc gas. Despite these limitations the result still provides useful information on these processes. We stress that the proposed coupling between onfall and radial flows together with the constraint of an exponential density profile now completely determines the spatial structure of onfall and radial flows. As the radial density profiles are kept fixed the mass distribution in each ring is not changed by the flows, but rather only by the continued accretion onto the galaxy. Hence, the radial flows only affect the distribution of the produced metals. Instead of using a preset formula for the flows throughout the disc these are 
derived consistently from the requirement of angular momentum conservation and the total required mass.

\section{Model behaviour}\label{sec:modelbe}
This part presents the results of the new prescription for coupling flows and onfall by the conservation of angular momentum. We start by reproducing the flow profile of the SB09 model to show that the mechanism works in principle. Then we discuss the basic dependency of the main observable constraint, the metallicity gradient, on the model parameters by showing illustrative examples.

\begin{figure}
 \centering
 \includegraphics[width=0.49 \textwidth,angle=0,keepaspectratio=true]{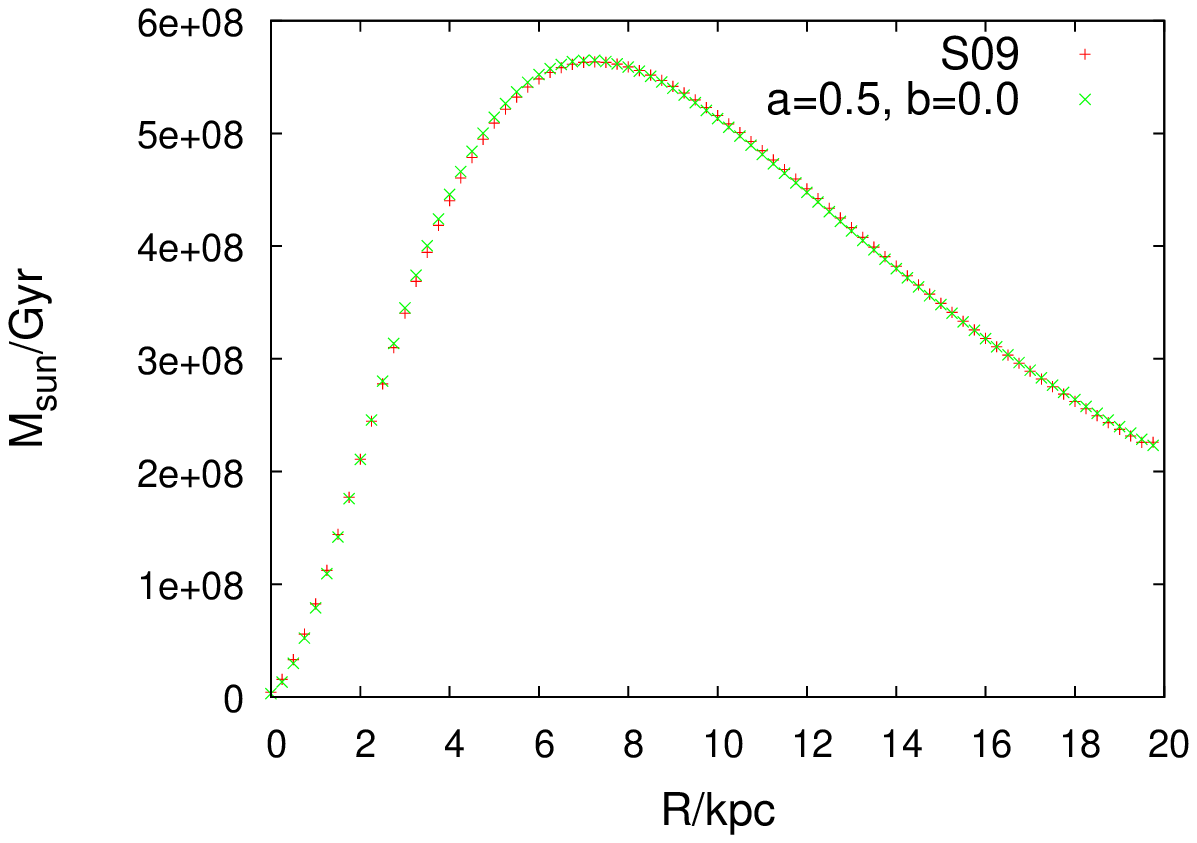}
 \includegraphics[width=0.49 \textwidth,angle=0,keepaspectratio=true]{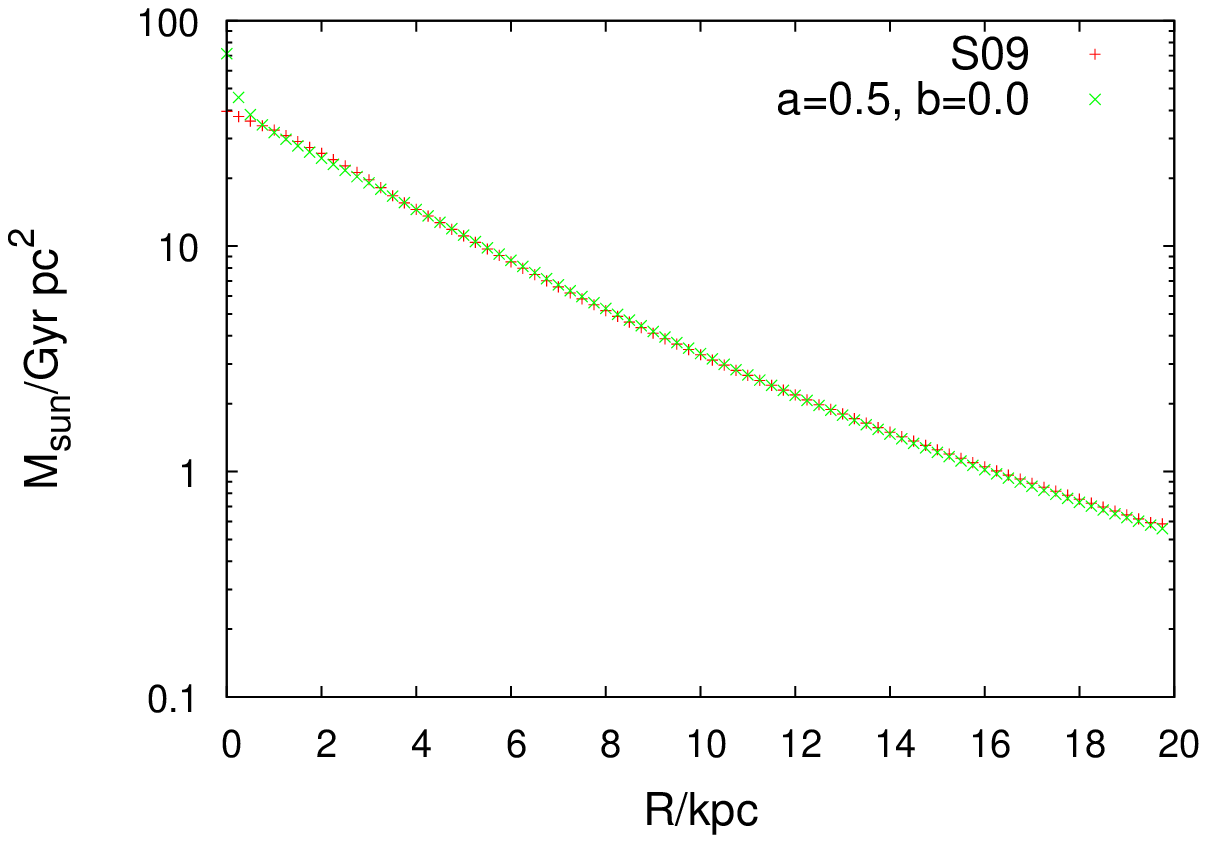}
  \caption{Upper panel: the rate of flow of gas over the annulus with radius R for the SB09-model and the new inflow model with $a=1/2$, $b=0$. Lower panel: the rates of accretion from the IGM per unit area of the disc.}
  \label{fig:referencecomparison}
\end{figure}

In Fig~\ref{fig:referencecomparison} the SB09 model is compared to the new scheme with parameters $a=1/2$ , $b=0$. All radial profiles are shown for the present day disc age of 11.7\eunit{Gyrs}. We see an acceptable agreement for this simple choice of parameters. This demonstrates on the one hand that the flows described by the naive formalism of SB09 were not unphysical and on the other hand we have an excellent starting point for our analysis by accurately reproducing the results of the older approach. A minor discrepancy resides in the innermost rings. It can be traced back to the huge relative angular momentum changes required for driving gas inwards in the central regions inhibiting proper radial flows, while the older approach simply set some fixed flow. Anyway this region is not of interest here due to a general lack of observations and the inadequacy of any model without a
reliable description of the Galactic bar.

\begin{figure}
 \centering
 \includegraphics[width=0.49 \textwidth,angle=0,keepaspectratio=true]{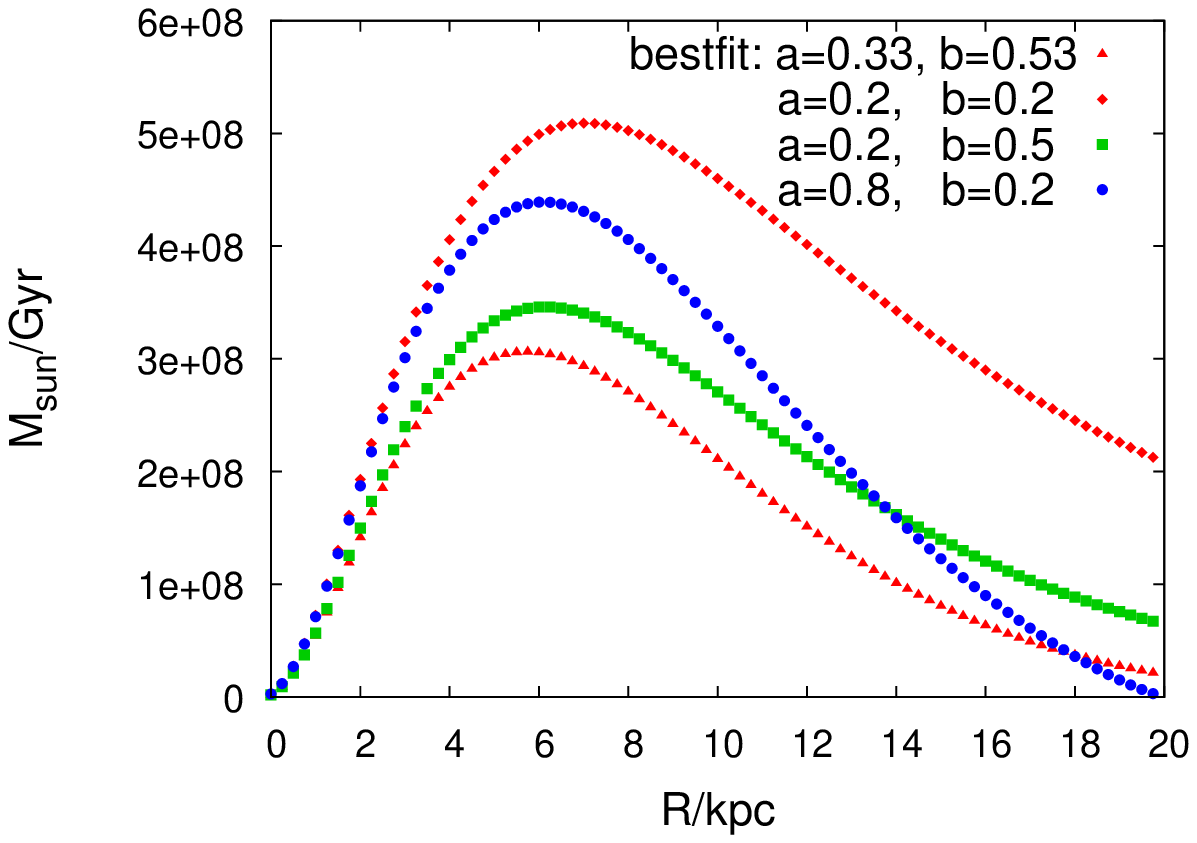}
 \includegraphics[width=0.49 \textwidth,angle=0,keepaspectratio=true]{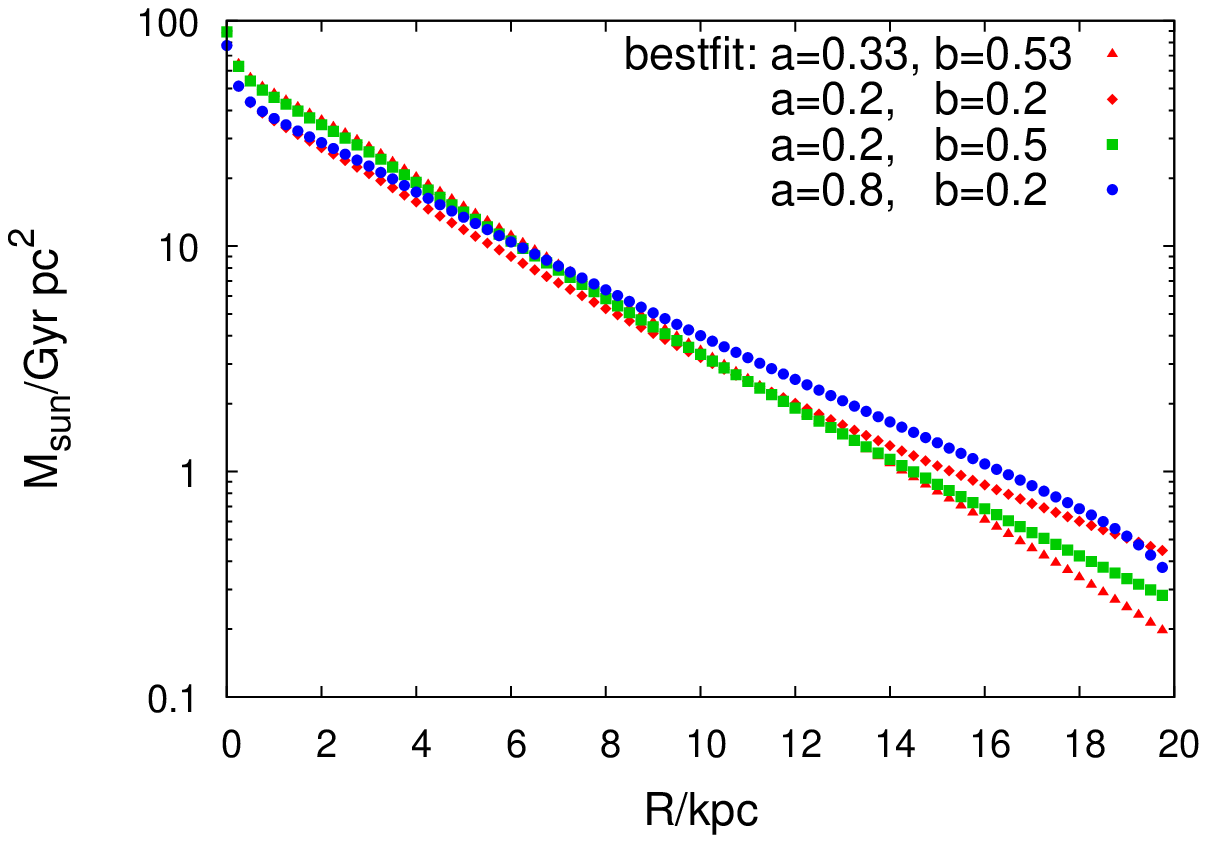}
  \caption{Upper panel: the rate of flow of gas over the annulus with radius R for new inflow model for the best fit model with $a=0.33$ for $b=0.53$, $a=0.2$ for $b=0.2$, $b=0.5$ and $b=0.7$ and $a=0.2$, $a=0.5$, $a=0.8$ for $b=0.2$. Lower panel: the rates of accretion from the IGM per unit area of the disc.}
  \label{fig:parametercomparison}
\end{figure}

We proceed to further study the behaviour of the flows for different parameter values. In Fig.~\ref{fig:parametercomparison} the resulting profiles for different choices of parameter $a$ and $b$ are shown.
The average angular velocity of onfalling material $\bar{v}= 0.5\,a+b$ will turn out to be an important quantity for the flows with direct consequences for the metallicity gradient of the simulations. For the models shown it is $\bar{v} = 0.7 , 0.3, 0.6$ and $0.6$, and we clearly see how the flows decrease with $\bar{v}$ increasing.

Thus, adjusting the parameter $b$ mainly sets the overall levels of inflow across the Galaxy as shown in the second (red diamonds) and third (green squares) model and it affects the whole disc in a similar fashion. The effects of parameter $a$ are strongest in the outer regions of the Galaxy as $a$ determines the angular momentum of the outer disc accretion and hence dominates the inflow behaviour there. Comparing the second (red diamonds) and fourth model (blue points) we see that increasing the parameter a leads to a stronger decline of the flows towards the outer regions of the Galaxy whereas the overall level is less affected than by changing $b$. In this way the radial extent of a high flow regime is controlled by setting the parameter $a$.
Finally it is interesting to consider the effects on the flows when varying a and b while keeping $\bar{v}$ fixed as for the third (green squares) and fourth (blue points) models. A higher $a$ in this case implies lower angular momentum in the inner regions which leads to a stronger buildup of the flows, and a higher maximal flow velocity. In the outer regions the angular momentum is actually higher, and the flows consequently decline. At fixed average angular velocity increasing $a$ increases radial flows in the centre of the Galaxy and decreases them in the outer regions.

In the lower part of \figref{fig:parametercomparison} we see the corresponding accretion densities. Albeit flattening at lower onflow angular momenta, which support the inner disc by radial gas flows, the accretion density is nearly exponential over most of the disc in all cases preserving the shape of the exponential surface density throughout the simulation.

\begin{figure}
 \centering
 \includegraphics[width=0.49 \textwidth,angle=0,keepaspectratio=true]{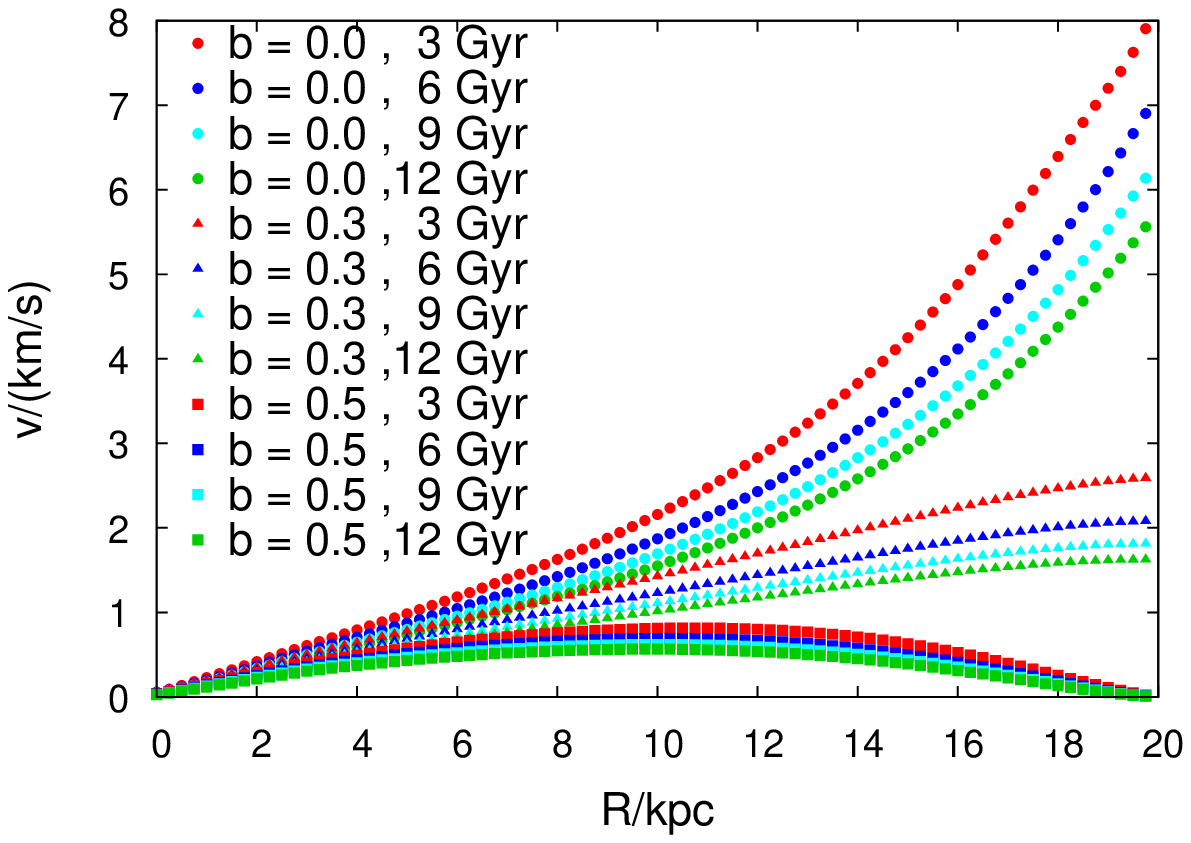}
 \includegraphics[width=0.49 \textwidth,angle=0,keepaspectratio=true]{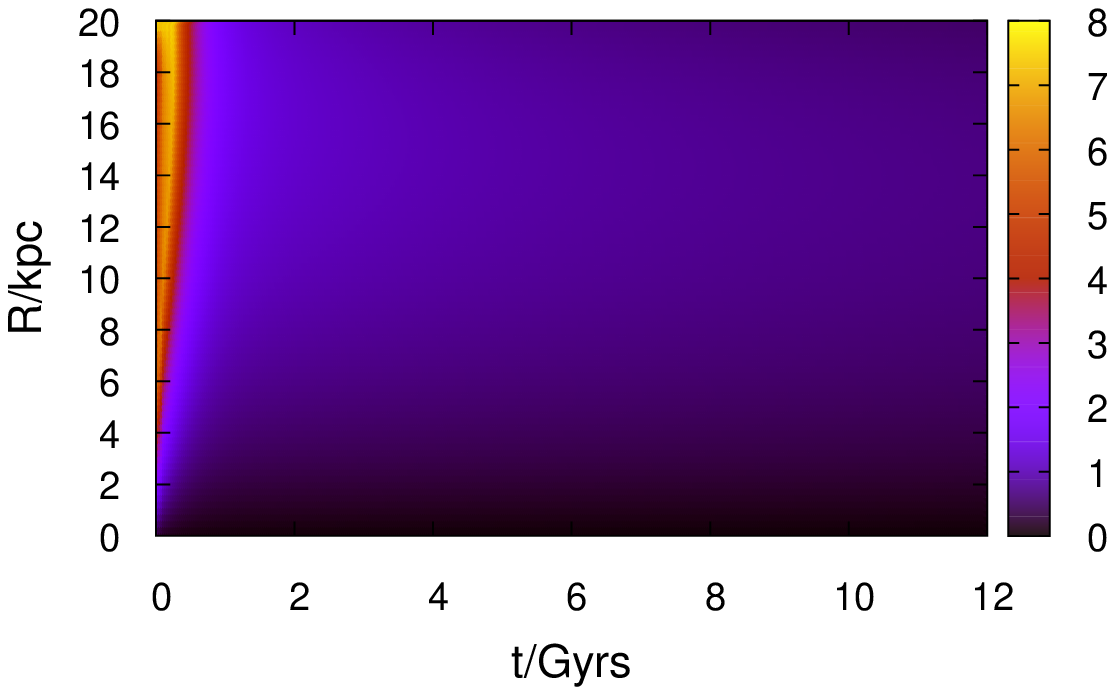}
  \caption{Upper panel: Resulting flow velocities for models with $a=0.5$ and $b=0.0,\, 0.3,\, 0.5$ at times 3, 6, 9 and 12\eunit{Gyrs}. Lower Panel: flow velocity map for the best fit model with $a=0.33$ and $b=0.53$ across the disc and the simulation time.}
  \label{fig:flowvelocities}
\end{figure}

As mentioned before, it should be checked whether the flow velocities stay within reasonable limits: we show this in the upper panel of Fig.~\ref{fig:flowvelocities} for different parameter values. During the buildup of the Galactic disc, little gas in the disc faces a fulminant onflow leading to high radial velocities as demonstrated by the streak of high velocities at the earliest times in the lower panel. During that period the flows are in some cases limited by the total mass content of the next outer ring. This issue is not overly important, since the behaviour at those early times does almost not affect the present day metallicity gradient. For parts of the parameter space (at low specific angular momentum) flow velocities can even get so high that they cannot be sustained at later times. Fortunately this is limited to otherwise uninteresting cases, where the onfalling gas has extremely low angular momentum. In that case this parametrisation fails and simulations with such parameter 
values should be discarded. For the parameter range of interest this behaviour is roughly limited to the first $\mathrm{Gyr}$ of the simulation as can be seen in the lower panel of Fig.~\ref{fig:flowvelocities}. Anyway the model currently does not provide a satisfactory description of early Galaxy formation.

The high flow velocities seen during the early times in the simulation should not be confused with an inside-out formation of the disc. Flow velocities increase outwards due to the build up of flows throughout the disc and the masses in each ring decreasing according to the exponential density profile.

\begin{figure}
 \centering
 \includegraphics[width=0.49 \textwidth,angle=0,keepaspectratio=true]{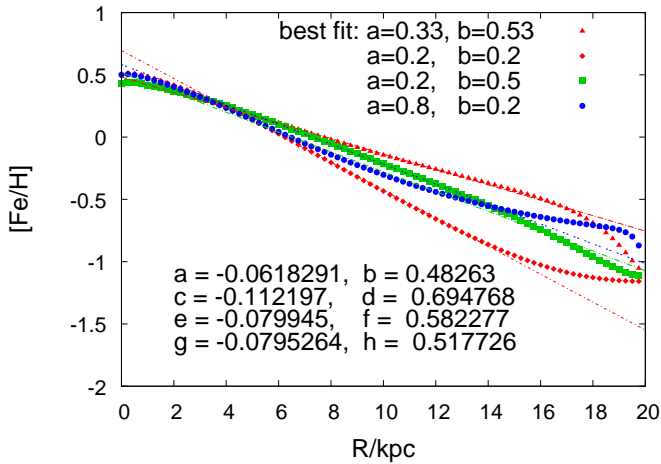}
  \caption{$\feh$-gradient for the models with different parameters. (gradient, y-axis-intersect) are (-0.112197, 0.694768), (-.079945, 0.582277), (-0.0795264, 0.0517726) and (-0.0618291,0.48262)}
  \label{fig:gradientscomparison}
\end{figure}

As our goal is to constrain the flows by fitting the Galactic radial abundance gradient we need to understand how the gradient depends on the flows. In Fig.~\ref{fig:gradientscomparison} the resulting $\feh$-gradient with linear fits in the region $ 4.0\eunit{kpc} < R < 14.6\eunit{kpc} $ is shown. As expected, an increase in onfall angular momentum, which lowers the flows, also lowers the metallicity gradient. In addition we clearly see the effect of depletion in the outer regions of the disc. If higher flows build up throughout the disc, the outer regions grab more gas from the IGM, which leads to a general decrease in metallicity and a stronger peak in the central regions, where the metals from stellar yields get advected. At the outermost ring the assumed metallicity of the fresh material dominates the local metallicity, since the model does not take into account any star formation beyond that point, so that the curves meet there.
The second (red diamonds) and third (green squares) models again illustrate the effect of parameter $b$ on the gradient. An overall lower flow level lowers the metallicity gradient. Moreover, lower overall flows limit the penetration depth of the depletion zone into the galactic disc which then leads to a sharper drop off in the outermost parts. The comparison between the second model (red diamonds) and the fourth model (blue points) shows the effects of changing parameter $a$. A higher value $a$ decreases the flows in the outer disc relative to the inner disc and hence causes the shown upwards bending of the metallicities.
Comparing model 3 (green squares) and 4 (blue points) which have the same average angular velocity, we observe that the gradients agree fairly well, at $-0.0799 \eunitfrac{dex}{kpc}$ and  $-0.0795 \eunitfrac{dex}{kpc}$ respectively, just the curvature is different. Consistent with the flow patterns discussed above, at fixed average angular velocity the model with higher $a$ has a higher metallicity and a steeper gradient in the very centre which flattens out towards the solar radius and the outer parts of the Galaxy.

Summarising the main points of this part we see that the model only works for sufficiently high angular velocity of the onfalling material corresponding to sufficiently low radial flows. However, flows exceeding our threshold of $8.2\eunitfrac{km}{s}$ produce unphysically large gradients and the metal advection to the central regions would require significantly larger mass loss of the central regions than currently assumed. The threshold is hence of no interest for the parameter region under study. The abundance gradient behaves as one naively expects with some complications due to the outer boundary. Fortunately, these boundary effects do not seriously affect the regions of the disc with $R < 17 \eunit{kpc}$.

\section{Fitting the model}\label{sec:Results}

\begin{figure}
\centering
 \includegraphics[width=0.49 \textwidth,angle=0,keepaspectratio=true]{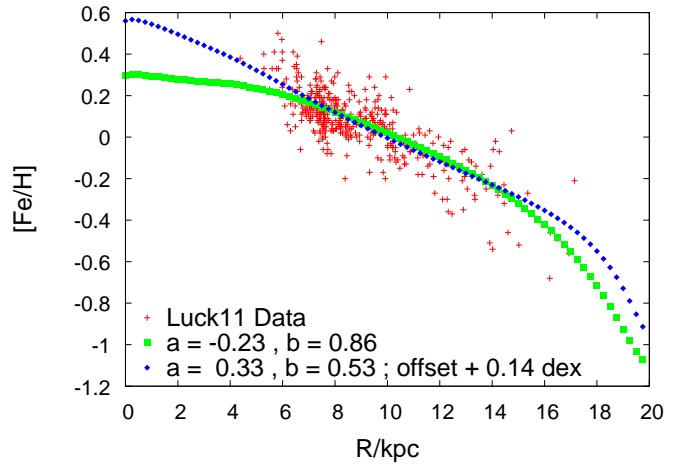}
\caption{$\feh$ from Cepheid Data as given in \protect\cite{Luck2011II} with both best fit models, the first obtained without an metallicity offset, the second with an offset by $+0.14$ dex.}
\label{fig:LuckData}
\end{figure}

In this section we will fit the model to data to the Galactic abundance gradient. We start by comparing the simulations with the complete catalogue of Cepheids given by \cite{Luck2011II} and references therein. They analysed a total of over 400 Cepheids and show that the $398$ stars passing their quality cuts deliver a linear relationship of metallicities with Galactocentric radius. $\feh(R)$ is best fit by $\feh = (-0.062 \pm 0.002) R/\kpc + (0.605 \pm 0.021)$. In Fig.~\ref{fig:LuckData} we see the observed data points and the best-fit models, the first fitted without and the second with a metallicity offset of $+0.14$ dex discussed below.

The parameter space of the new flow model is explored on a grid consisting of $992$ simulations equispaced on the $(a+b)$-$b$-parameter plane. When interpreting the following plots it is useful to keep in mind that the case of constant angular velocity ($a=0$) corresponds to the diagonal from the lower left to the upper right corner. To the left and over this line $a<0$ while to the right and below it $a>0$. Lines of constant average angular velocity $\bar{v}= 0.5\,a+b$ correspond to parallels to the diagonal from the upper left to the lower right corner.

To assess the agreement between the data set and the simulations we computed
\begin{equation*}
 {\chi}_{red}^{2} = \frac{1}{N-2}\sum_{i=1}^{N} \frac{\left[(y_i - y_{i,sim})\right]^{2} }{{\sigma}^{2}}
\end{equation*}
where $y_{i,sim}$ is the [Fe/H] value of the simulation at radius $x_i$.

\begin{figure}
\centering
 \includegraphics[width=0.49 \textwidth,angle=0,keepaspectratio=true]{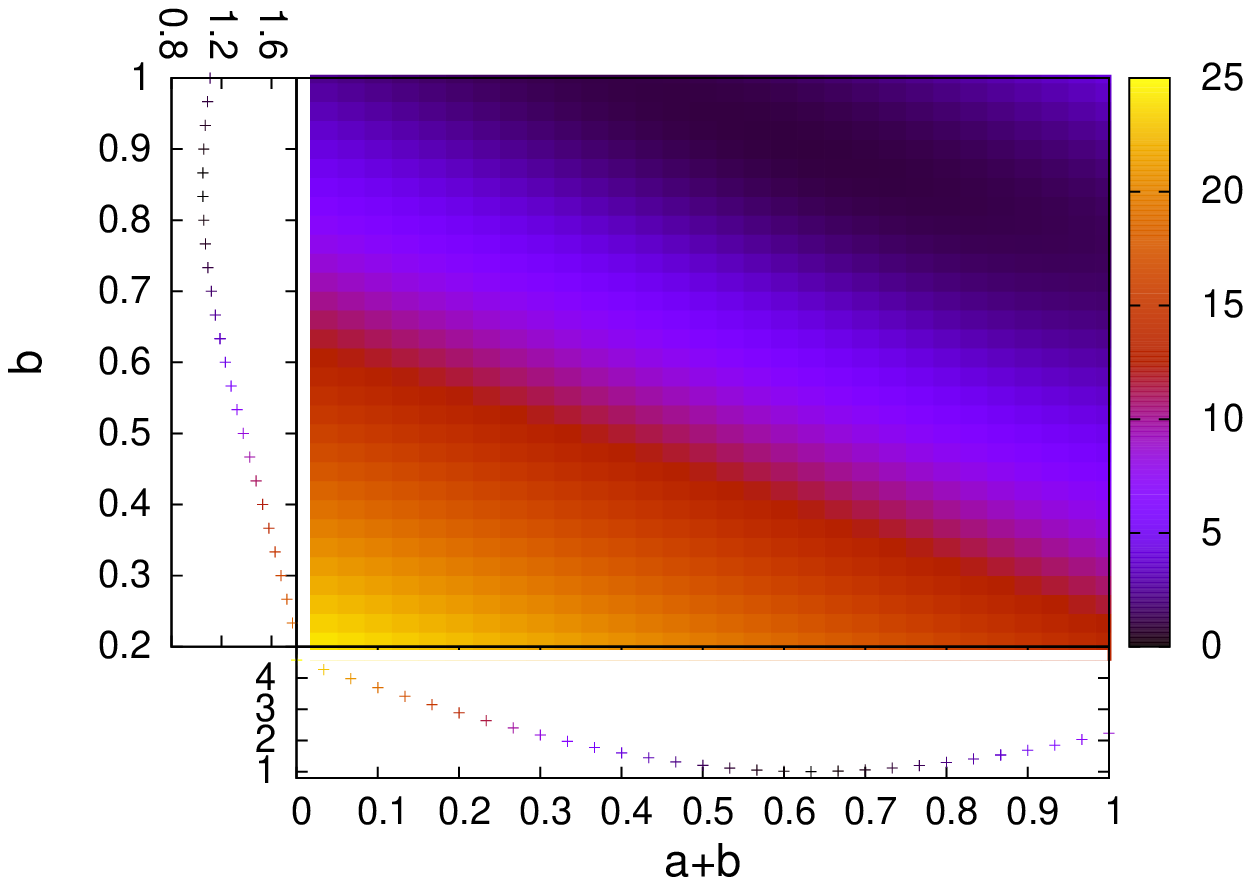}
 \includegraphics[width=0.49 \textwidth,angle=0,keepaspectratio=true]{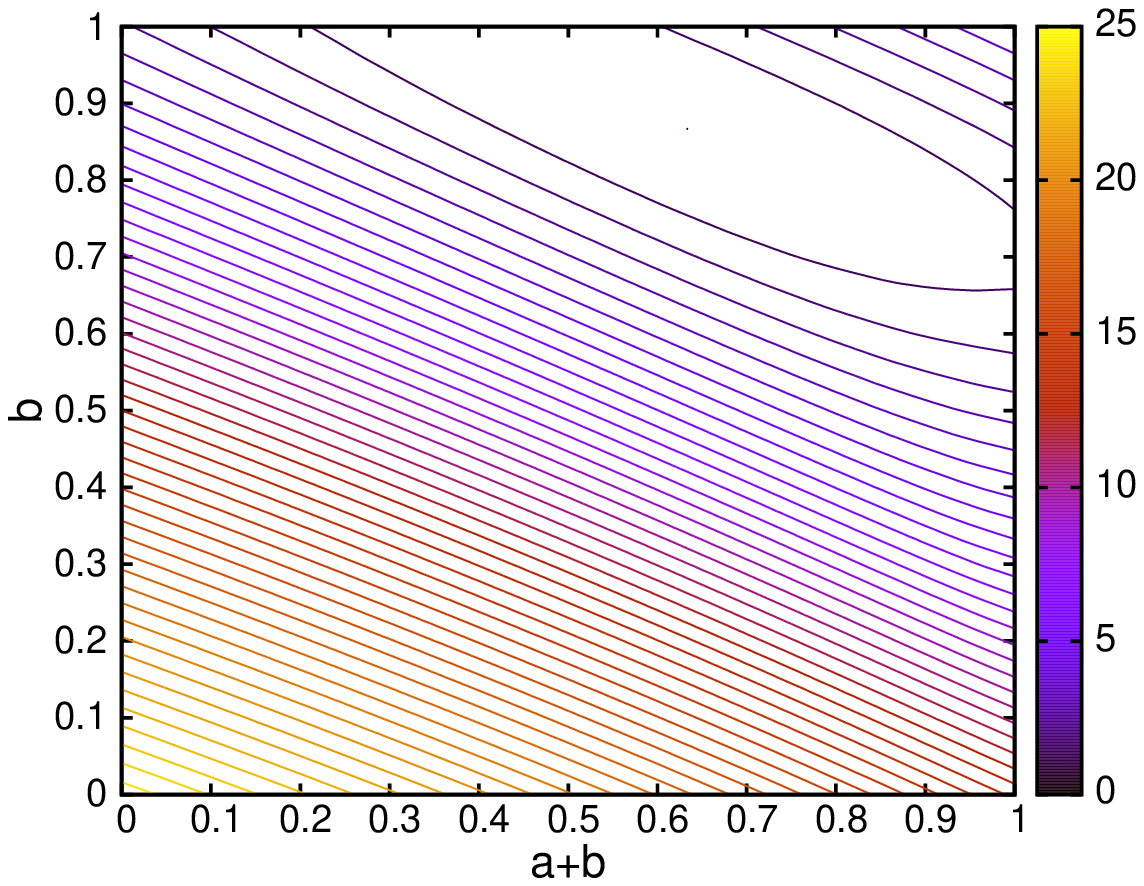}
\caption{Upper panel: ${\chi}_{red}^{2}$ for comparison with Luck11II Data on the simulation grid(main plot), $a+b=0.6333$ -slice (left of main plot) and $b=0.8666$ -slice (bottom of main plot). Lower Panel: ${\chi}_{red}^{2}$-contours equispaced by 0.5 increments between 1 and 25.}
\label{fig:chimap_11_II}
\end{figure}

In Fig.~\ref{fig:chimap_11_II} we plot ${\chi_{red}}^{2}$-values for the entire grid and parameter slices intersecting at the absolute minimum are displayed on the rim. Contours are shown on the lower panel. By looking at the contour plot the unique minimum at $a=- 0.233$, $b=0.8666 $ can be identified. This corresponds to an average angular velocity of the onfalling material of $v/\Vc = 0.75$ or $v=165\eunitfrac{km}{s}$ assuming $\Vc=220\eunitfrac{km}{s}$.

With this naive approach the fit results depend strongly on the detailed assumptions about chemical evolution, because a change in the total metallicity will shift the minimum and alter its depth in our parameter space. The absolute metallicity is e.g. impacted by the specific stellar yield dataset in use or by a different IMF, which changes the ratio between lock-up and yields, or by the overall loss rate in Galactic winds, factors which are quite weakly determined. Aditionally the determination of stellar abundances may suffer from systematic errors of the order of $0.1\, \mathrm{dex}$. Especially the Salpeter IMF has the largest mass of the known IMFs in use, which leads to a probable lock-up of material in low-mass stars, which in turn lowers absolute metallicity levels. To the contrary, relative abundances are known to high accuracy and the gradient in the simulations is far more robust as well.

So for this purpose it is far more reliable to treat the average metallicity as a free parameter. This still gives a unique best model and improves the fit for parameters away from this minimum. The quality of the best fit is comparable for both procedures as they differ mostly in the inner parts of the Galaxy where no data constrains the models. The best fit with an allowed offset shows a linear gradient across almost the whole Galaxy, whereas the model fitted without the offset flattens towards the centre. Effectively, allowing for an offset fits the form of the gradient resulting in linear abundance gradients over the whole disc for the best fit model.

\begin{figure}
\centering
 \includegraphics[width=0.49 \textwidth,angle=0,keepaspectratio=true]{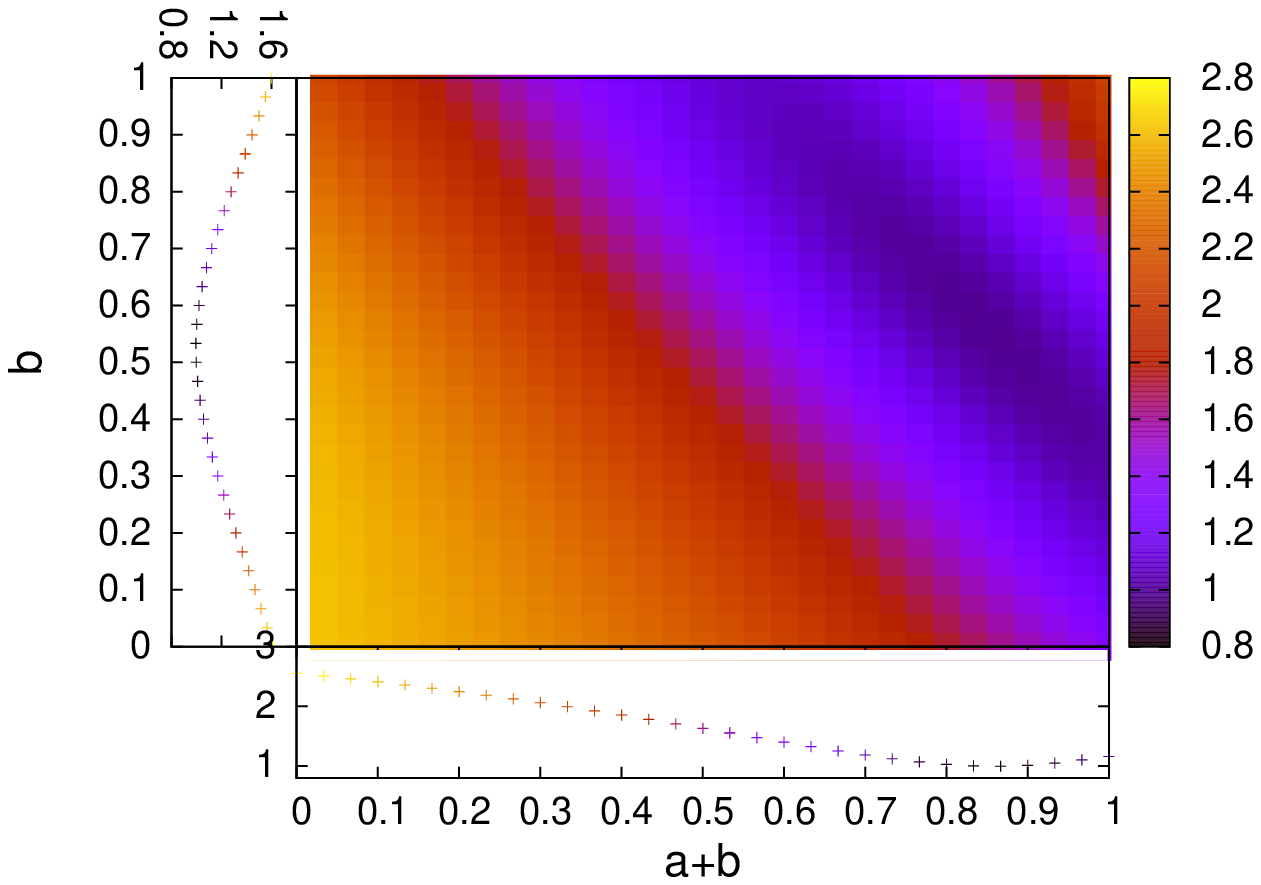}
 \includegraphics[width=0.49 \textwidth,angle=0,keepaspectratio=true]{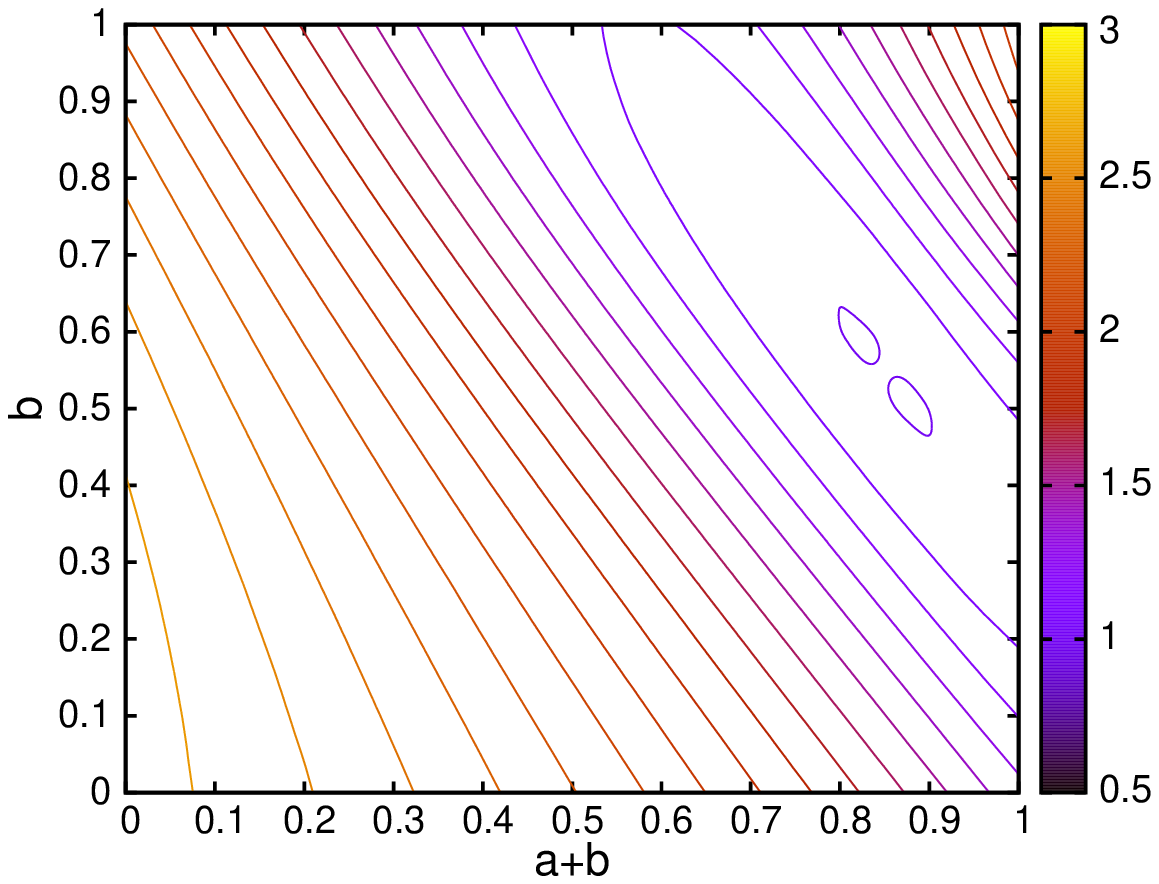}
\caption{Upper panel: ${\chi}_{red}^{2}$ for comparison with Luck11II Data on the simulation grid(main plot), $a+b=0.0.8666$ -slice (left of main plot) and $b=0.5333$ -slice (bottom of main plot). Lower Panel: ${\chi}_{red}^{2}$-contours equispaced by 0.1 increments between 1 and 3}
\label{fig:fit_chimap_11_II}
\end{figure}

\begin{figure}
\centering
 \includegraphics[width=0.49 \textwidth,angle=0,keepaspectratio=true]{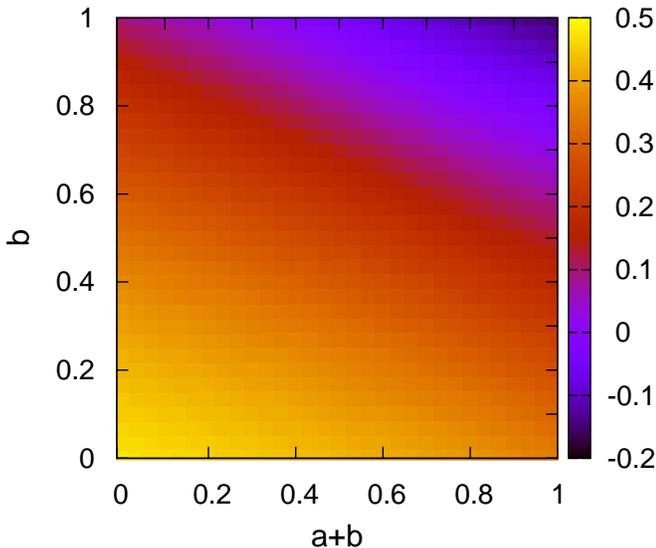}
\caption{Offset used for fitting the simulations to the Luck11II Data on the simulation grid}
\label{fig:11_II_offset}
\end{figure}

Fig.~\ref{fig:fit_chimap_11_II} shows the corresponding results when fitting the average metallicity of the simulations to the data. The offsets used for the best fits are shown in Fig.~\ref{fig:11_II_offset}. Generally higher flow velocities decrease the average metallicity in the simulations, but the offset is relatively small in the region where the best fits are achieved. The absolute minimum is now at $a= 0.3333$, $b=0.5333 $ with an offset of $0.14 \, \mathrm{dex}$, the simulations having the lower average metallicity. The average angular velocity is $v/\Vc = 0.7$ or $v=154\eunitfrac{km}{s}$.

\begin{figure}
\centering
 \includegraphics[width=0.49 \textwidth,angle=0,keepaspectratio=true]{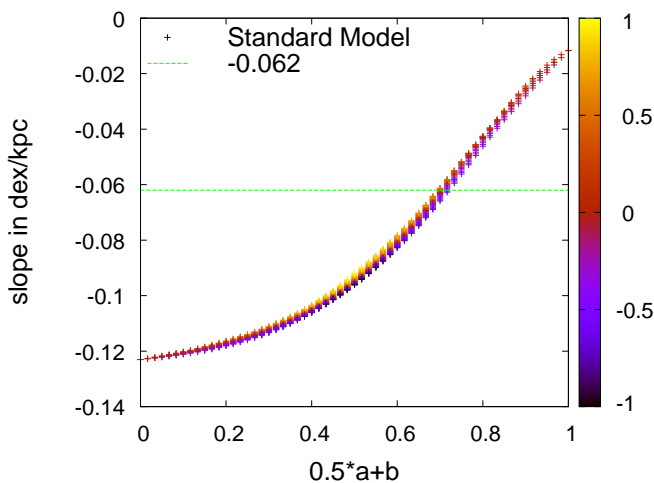}
\caption{Metallicity gradient of the simulations as a function of the average angular velocity of the onfalling gas, the dependence on parameter a is colour-coded.}
\label{fig:completegrid_slope}
\end{figure}

Finally we note that the metallicity gradient of the simulations is tightly correlated with the average angular velocity $\bar{v}$ of the onfalling gas as seen in Fig.~\ref{fig:completegrid_slope}. This relation grows worse when using a different value, e.g. the flow velocity at the solar radius, instead of the average. Thus, the results lead us to consider the average angular velocity as an important quantity to understand the simulations. As discussed in the preceding section at fixed average angular momentum the models with higher $a$ tend to have a slightly flattened metallicity gradient. This creates some scatter around the average relation depending on the precise values of the parameters $a$ and $b$ which justifies fitting the models on the whole grid instead of only working with the average velocities. But the qualitative behaviour of the simulations can clearly be extracted from this plot alone and the best fit can be determined up to an uncertainty of $\Delta
v/\Vc = 0.07$.

In summary, the average velocity can be constrained to be $0.70\,\Vc  = 154\eunitfrac{km}{s} \le v \le 165 \eunitfrac{km}{s} = 0.75 \, \Vc$ . The average velocity would be a plausible value for gas from the corona. The corona presumably rotates slower than the disc gas by about $v_{lag,cor} \approx 75 \eunitfrac{km}{s}$, which corresponds to $v_{cor} \approx 145 \eunitfrac{km}{s}$ when neglecting any vertical velocity gradients \citep{Marasco2011,Marinacci2011}. Apparently, this result depends on the gradient to be fitted and we cannot claim that this result would be unique as there are different ways of producing a gradient. Still, there seems to be a strong tendency to recover such values within this framework.

\subsection{Additional inflow}

As discussed before, both (turbulent) friction in the gas layers and interaction of the gas with spirals and bars can drive a radial inflow which could be of the order of 0.3\eunitfrac{km}{s}. To ascertain how strong this affects our results, we computed simulations in which a constant radial inflow velocity was added to the inflow model.
Because this simply increases the overall flow levels, we generally expect the gradient to steepen at the same average angular velocity $\bar{v}$ of the onfalling material. Hence, the best fit should shift to higher values of $\bar{v}$ reducing the flows caused by onfalling material. However, the nature of these flows is drastically different from those caused by the onfalling material, which is not constant, but varies radially due to the coupling to the infall.

\begin{figure}
\centering
 \includegraphics[width=0.49 \textwidth,angle=0,keepaspectratio=true]{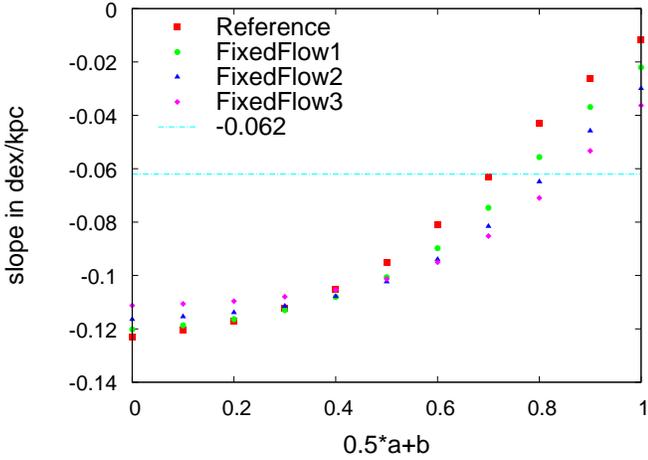}
\caption{Metallicity gradient of the simulations as a function of the average angular velocity of the onfalling gas, for the reference simulation and with an additional inflow with constant speed of $v=$ 0.2, 0.4 and 0.6\eunit{km/s} for models 1,2 and 3 respectively}
\label{fig:fixedflow_slope}
\end{figure}

Specifically, we add an additional constant inflow with a speed of $v=$ 0.2, 0.4 and 0.6\eunit{km/s} to the simulations. Fig~\ref{fig:fixedflow_slope} shows the resulting metallicity gradients for the simulations. We see that the additional flow has a significant effect over the whole parameter range. For $\bar{v} \ge 0.4$ the simulations with the additional flows show a considerably steeper gradient, lying below the reference values (red squares).
Below $\bar{v} = 0.4$ the simulations show a shallower gradient being above the reference values. However, this region is not important as the simulations there show an unreasonably high gradient that is not observed. Moreover, the flows in this regime are so strong that they are actually limited by the maximally allowed flows of the simulation. For these parameter values the flows saturate in the centre of the galaxy concentrating most of the produced metals in the innermost ring and the strong inflow across the whole disc transports low metallicity gas from the outskirts of the Galaxy towards the inner regions reducing the average metallicity and gradient there. Thus, we disregard the models in this parameter regime.

In the region of interest, the differences roughly amount to $\Delta v / \Vc = +0.07$ for the simulation with an additional constant flow speed of 0.2\eunit{km/s}, and the differences are correspondingly higher for the other simulations at $\Delta v / \Vc = +0.11$ and $\Delta v / \Vc = +0.15$ for 0.4\eunit{km/s} and 0.6\eunit{km/s} respectively.

We conclude that additional radial flows across the whole Galactic Disc strongly influence the abundance gradients. But the flows due to onfall remain important for the gradient even in this case. The models considered here are extreme cases in the sense that they assume an additional constant inflow velocity that extends across the whole disc and consequently the effects on the gradient are relatively strong. Moreover, only the first model is actually within the estimates on the order of the flows caused by the considered effects, and the second and third model already have higher than expected flow levels.

\subsection{Effects of a different infall history}
The prescription for the infall used in the original model might also influence the results obtained from the flow model. In particular, the evolution of the infall determines the available amount of gas and the SFR history. With a non-linear SFR law increasing the mass directly creates a gradient due to the stronger metall production in the centre of the Galaxy. As in our model the radially flowing mass is a fraction of the infalling mass, lower infall also reduces the flow levels across the whole disc yielding lower metallicity gradients. In addition a lower gas mass lowers the maximal gas flows, which may lead to differences from the expected behaviour for very high flow velocities, and increases the effects of flows on the gradient, an effect studied below.

In the standard model infall increases the gas mass in the disc according to
\begin{equation*}
 \dot{M}=\frac{M_1}{b_1}\, e^{-t/b_1} + \frac{M_2}{b_2} e^{-t/b_2}
\end{equation*}
with the parameters $b_1=0.3 \eunit{Gyrs}$ and $b_2=14 \eunit{Gyrs}$. To test the dependence on the assumed infall history the parameter $b_2$ of the second exponential was changed to $b_2=7 \eunit{Gyrs}$ and $b_2=28 \eunit{Gyrs}$ corresponding to a faster or slower decline of infall respectively. Differences between the simulations should be higher for the case of the faster decline because the original simulation already has a relatively slow decline and the exponential decay law.

\begin{figure}
 \centering
 \includegraphics[width=0.49 \textwidth,angle=0,keepaspectratio=true]{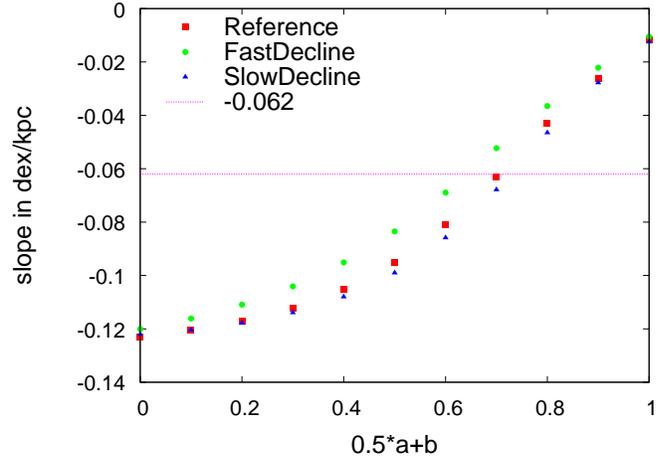}
 \caption{Metallicity gradient of the simulations as a function of the average angular velocity of the onfalling gas for the simulations with a fast or slow decline of the infall and the standard model}
 \label{fig:infalltest_metallicitygradient}
\end{figure}

In Fig.~\ref{fig:infalltest_metallicitygradient} the results of the simulations are shown as a function of the average angular velocity.
The simulations with the faster decline shows a considerably lower metallicity gradient for almost all parameter values in agreement with their lower flow levels, lying above the standard model in the plot, with a maximal difference up to $0.02\eunit{dex/kpc}$. For decreasing $\bar{v}$, thus, higher flows, the difference to the standard model gets smaller. In this parameter regime, the standard model's flows start to be limited by the total available gas mass, while this is not yet the case for the generally lower flows of this model.

The simulations with a slower decline show metallicity gradients comparable to the standard model, but with a slightly higher gradient for most of the parameter values corresponding to generally higher flows. For the very highest flow velocities, the gradient is actually lower. However, this difference is quite small and may be explained by saturation of the flows in the inner region of the galaxy and higher flows in the outer parts reducing the average metallicity.

Regarding the best fit to the observed data, these differences would translate at most into $\Delta v / \Vc = 0.1$ in the region of interest.

A more detailed analysis can again be performed on a grid of the parameter values, in the case of the fast decline model the best fit values are $a= 0.3$, $b=0.5$ corresponding to $v/\Vc = 0.65$, for the slow decline we find $a= 0.366$, $b=0.533$ corresponding to $v/\Vc = 0.716$, both quite close to the standard model values of $a= 0.333$, $b=0.533 $ and $v/\Vc = 0.7$.

In summary doubling the decay time has a small effect of $\Delta v / \Vc = +0.016$, whereas halving it leads to a difference of $\Delta v / \Vc = -0.05$ compared to the standard model.

\subsection{Effects of different star formation rates}
A different star formation efficiency has an effect on the flow model by changing the amount of gas available in the disc and on the metallicity by changing the SFR history. The flow model is affected as changing the gas mass in the disc changes the fraction of onfalling to disc gas and the total angular momentum in the disc. For higher gas masses in the disc the effect of flows are reduced as the same amount of metals transported radially by flows is diluted in a higher amount of disc gas.
Additionally, the simulations will show different behaviour in the high flow regime when the flows are limited by the total mass in the rings.

In the standard model stars form according to
\begin{equation*}
\frac{d{\Sigma}_{star}}{dt}= k\, {\Sigma}^{1.4}_{g}\, .
\end{equation*}
Changing the efficiency of star formation ($k$) also changes the stellar mass in the disc and thus the metals produced in stars. To avoid this problem and to isolate the effect on the flow model we kept the star formation history of the simulations constant by also using different values for the masses $M_1$ and $M_2$ in the infall law. When increasing the star formation efficiency, the available gas mass was lowered and vice versa. Thus, mainly the gas mass in the disc is affected. However, the same amount of metals produced is then diluted in more or less gas, also changing the gradient.

The standard model with $k=1.2 \times 10^{-4}$, $M_1 = 0.45\times 10^{10} \msun $, $M_2= 2.9 \times 10^{10} \msun$ was compared to simulations with a higher efficiency and lower gas mass $k=3.167\times 10^{-4}$, $M_1 =  0.11\times 10^{10} \msun $, $M_2= 3.07\times 10^{10} \msun$ as well as simulations with a lower efficiency and higher gas mass $k=0.3789\times 10^{-4}$, $M_1 = 1.5\times 10^{10} \msun $, $M_2= 2.50 \times 10^{10} \msun$.

\begin{figure}
\centering
 \includegraphics[width=0.49 \textwidth,angle=0,keepaspectratio=true]{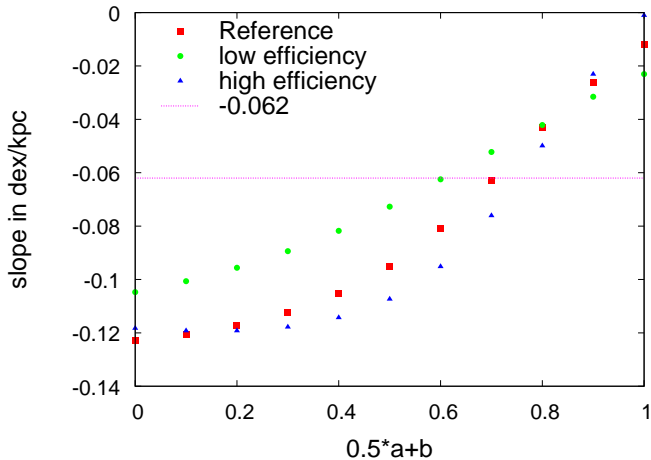}
\caption{Metallicity gradient of the simulations as a function of the average angular velocity of the onfalling gas for simulations with different star formation efficiency}
\label{fig:Kennicutt_slope}
\end{figure}

We expect the simulations with the higher gas masses to have a smaller metallicity gradient because higher gas masses should decrease the radial flow velocities of the gas at fixed $\bar{v}$. This is generally confirmed in Fig.~\ref{fig:Kennicutt_slope} which shows the resulting gradients in the simulations.
The simulations with the lower efficiency and higher gas mass show a shallower gradient over most of the parameter values. But for very low flow velocities, the gradient is steeper than in the standard model. Here the effects of a higher gas mass dominate over the effects of the flows and steepen the gradient. This is best seen for $\bar{v} = 1.0$, which corresponds to no flows, for which the low efficiency high gas mass simulation shows the strongest gradient, the standard simulation a somewhat smaller gradient and the high efficiency low gas mass simulation a nearly flat gradient.

For the high efficiency simulation three regimes of parameter space in which the simulations show different behaviour compared to the standard model can be identified. The discussed low flow regime, where the effects of gas mass and dilution dominate. The intermediate regime, where the direct effects on the flow model become visible, and the simulation has a steeper gradient corresponding to it's lower gas mass. And the very high flow regime, in which the gradient flattens slightly because the flows saturate in the centre of the galaxy and transport more low metallicity gas into the inner regions lowering the average metallicities.

Again the differences in the region of interest are relatively small and give roughly $\Delta v/\Vc = 0.1$, the lower efficiency simulation yielding a lower average velocity and the higher efficiency simulation giving higher values.

We conclude that different assumptions for the star formation rate do influence our results moderately by changing the gas mass in the disc. Fortunately, the effects are relatively small in the region of the observed metallicity gradient giving an uncertainty of $\Delta v/\Vc = 0.1$ for the simulations presented here.

\section{The accretion profile}\label{sec:accprofile}

While the angular momentum is in formidable agreement, \cite{Marasco2011} also predict a specific accretion profile for their theory of supernova-driven gas accretion which greatly differs from that seen in our models.

To test whether their accretion profile allows for a plausible chemical evolution, we computed simulations based on the SB09 model. Assuming the accretion profile, we normalise the total accreted mass in each timestep to be the need of the whole disc and distribute the infall over the annuli according to the profile. The remaining missing mass in each annulus is then provided for by radial flows.

\begin{figure}
 \centering
 \includegraphics[width=0.45 \textwidth,angle=0,keepaspectratio=true]{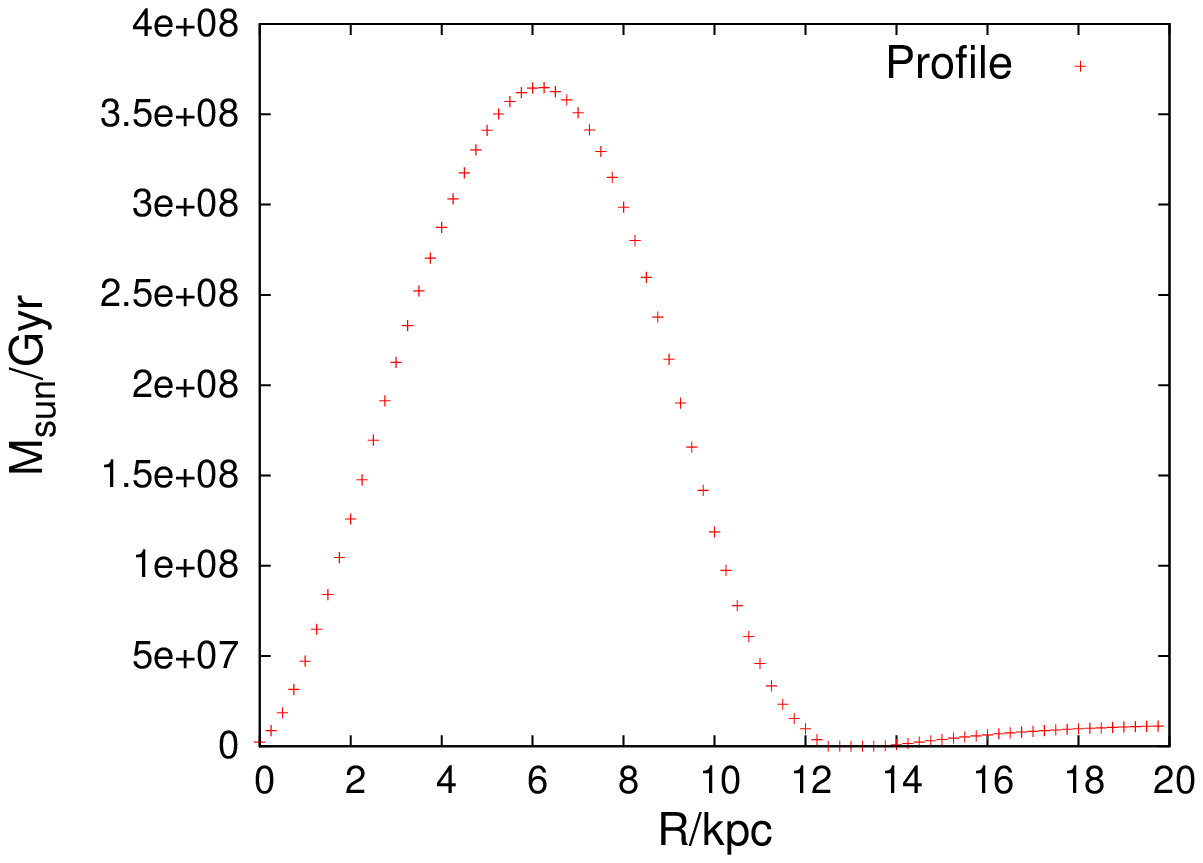}
 \includegraphics[width=0.45 \textwidth,angle=0,keepaspectratio=true]{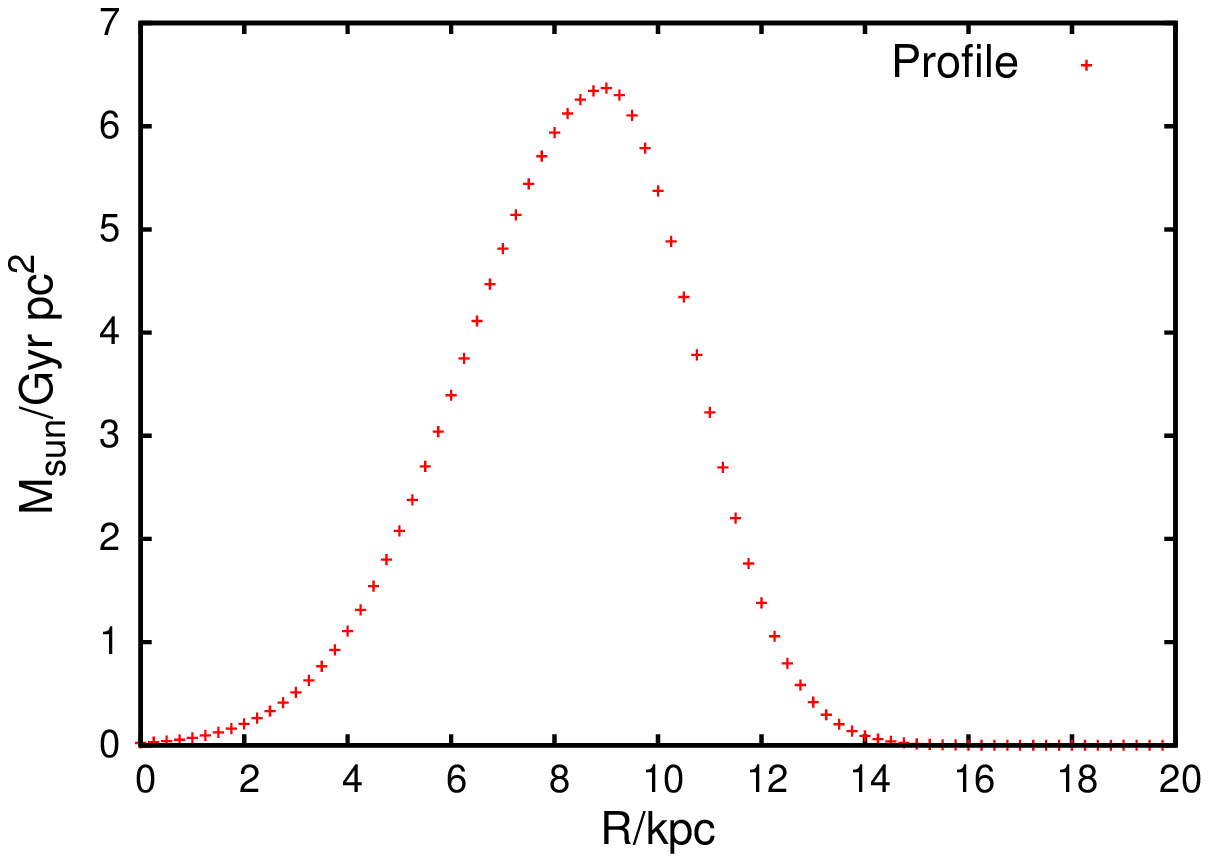}
 \includegraphics[width=0.45 \textwidth,angle=0,keepaspectratio=true]{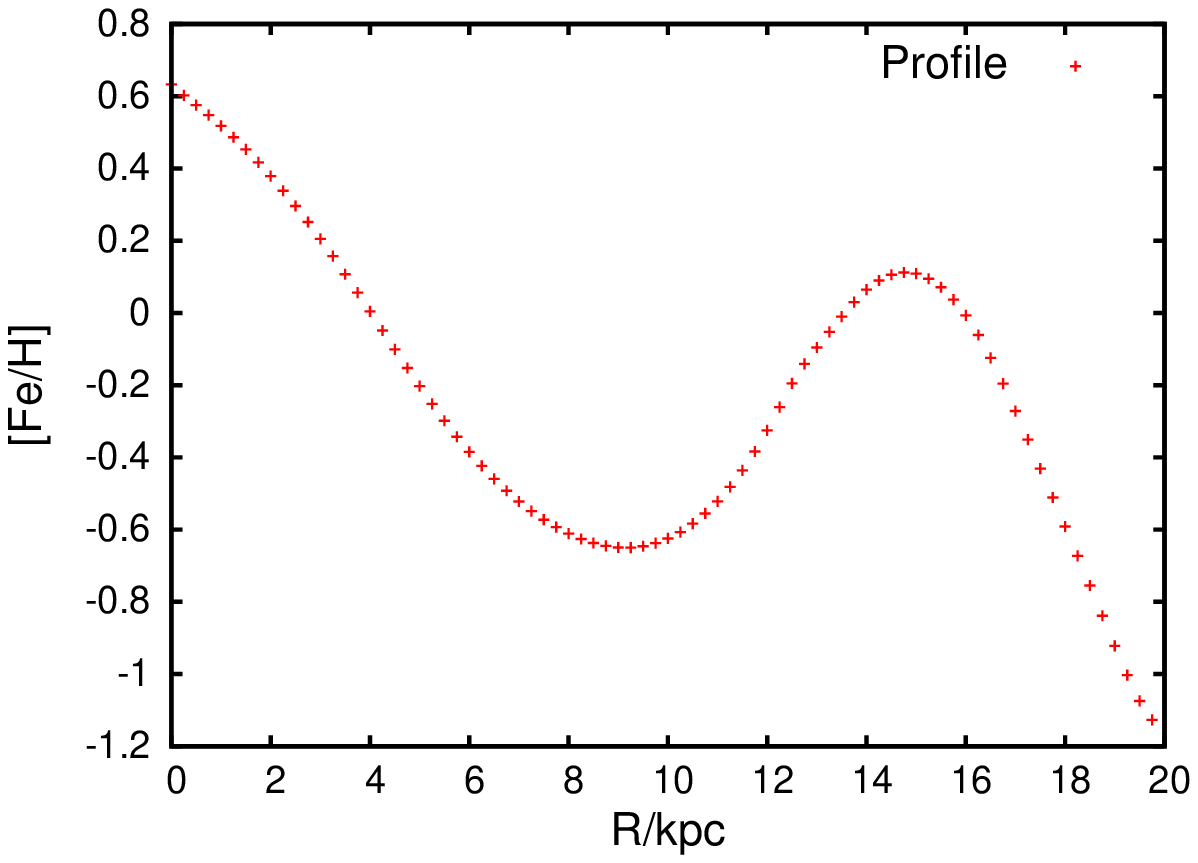}
 \caption{Upper panel: the rate of flow of gas over the annulus with radius R for the SB09-model with the accretion profile. Middle panel: the rates of accretion from the IGM per unit area of the disc. Lower Panel: Abundance gradient assuming the accretion profile of Marasco et al. (2011)}
 \label{fig:profile}
\end{figure}

Results of this model are shown in Fig.~\ref{fig:profile}. As the peak of the accretion rate is farther out than the peak of the star formation rate, the inner regions of the disc need to sustain their mass by high radial inflows. These high flows cause a steep gradient to build up in the inner half of the disc ($R< 9 \eunit{kpc}$). As the infall again falls off beyond $9\eunit{kpc}$ more of the need in each ring is provided by radial inflows, but without the diluting effect of the low-metallicity infall from the IGM the metallicity increases between 10\eunit{kpc} and 15\eunit{kpc}. Beyond that radius both the low star formation rate and the inflow from the outermost rings serve to decrease the metallicity in the gas.

There are two main shortcomings of the resulting abundance gradient. First of all the very steep abundance gradient of  $-0.164 \eunitfrac{dex}{kpc}$ in the region $R< 9 \eunit{kpc}$. This is an immediate consequence of the high radial flows necessary to redistribute the infalling gas inwards to the maximum of the star formation rate in the Galaxy. If such an accretion profile without additional infall from a different source is adopted this effect is inevitable and strongly disagrees with the gradient from observational data. Eventhough \cite{Luck2011II} find a somewhat steeper gradient of $-0.103\eunitfrac{dex}{kpc}$ in the inner disc $R <6.6 \eunit{kpc}$, this still differs greatly from the model results. Secondly the increase of the metallicity between 10\eunit{kpc} and 15\eunit{kpc} is definitely ruled out by the data. However, it might be easier to remove this problem by adding a different source of infall that provides enough gas in the outer rings to reduce the metallicity but does not greatly
affect
the global accretion rate due to the small masses required. In the inner disc it would be difficult to solve this problem, because we are facing an actual mass deficit and not a mere gradient problem, so we cannot simply blame it on the bar. The most likely way out is that the assumption of \cite{Marasco2011} of a minimum altitude above the disc that a gas cloud has to reach in order to accrete coronal gas fails or that multiple supernovae can assemble more gas there.

\section{Non-circular orbits and angular momentum transport}\label{sec:dyingstars}
As mentioned in the introduction angular momentum might also be transported by stars moving on non-circular orbits. We assume that a star is born at some radial position with the corresponding angular momentum of the gas. During its lifetime it may leave this radial region and eject material at a different radial position. Depending on whether it spends more time outside or inside its original radius this will cause inflows or outflows. Two processes that are included in the SB09 model are important for this process: churning and blurring. If one uses a simplified picture of stellar orbits showing radial oscillations around their guiding centre radius, churning changes the guiding centre radius and the effects of the radial oscillation are described by blurring. Churning exchanges stars and gas between different annuli by means of a resonant interaction with transient spiral arms in the Galaxy. In this interaction total angular momentum is conserved, the star actually lowers (when moving inwards) or
increases (when moving outwards) its angular momentum. Flows associated with the spiral pattern are hence driven by the different radial profile of gas and stars, which are already in the model, and the gas losing a bit of angular momentum in shocks in the spiral arms, which is generally a minor effect.

Blurring affects stars that do not move on circular orbits and, hence, spend time at galactocentric radii away from their guiding centre radius. If a star ejects material at such a point, its yields have a different specific angular momentum than the surrounding gas, which will cause in- or outflows. Generally a larger dispersion makes populations expand outwards. This process depends on the shape of the potential and the radial density and velocity dispersion profiles, but in general there is a positive asymmetric drift that drives an inflow.

The model with parameters $a = 0.5$, $b=0$ is used to estimate this effect because this choice closely resembles the original SB09 model. At this point the resulting flows are only calculated and not actually implemented. This should not have any impact on our conclusions in this part as the mass distribution is held fixed by the onfall mechanism. For the same reason the model parameters $a$ and $b$ have no effect on the velocity profiles and are only given for completeness. If the flow velocities are comparable to those resulting from the onfalling gas, it would certainly change the chemical composition of the disc and need to be considered.

\begin{figure}
 \centering
\includegraphics[width=0.49 \textwidth,angle=0,keepaspectratio=true]{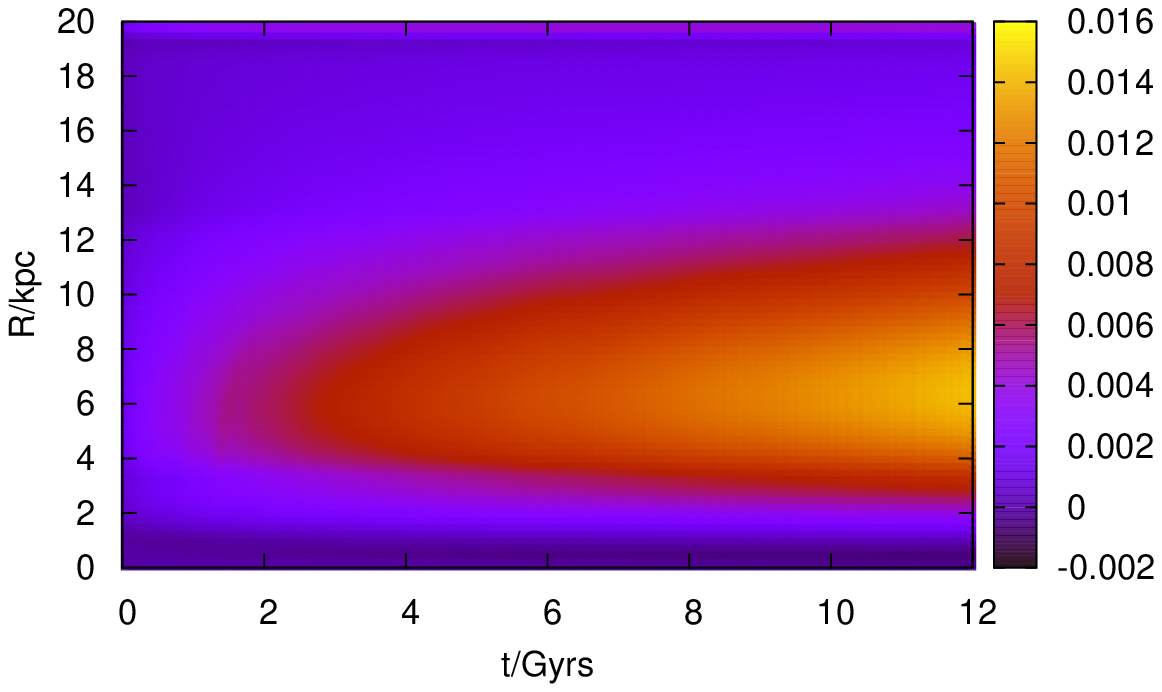}
\includegraphics[width=0.49 \textwidth,angle=0,keepaspectratio=true]{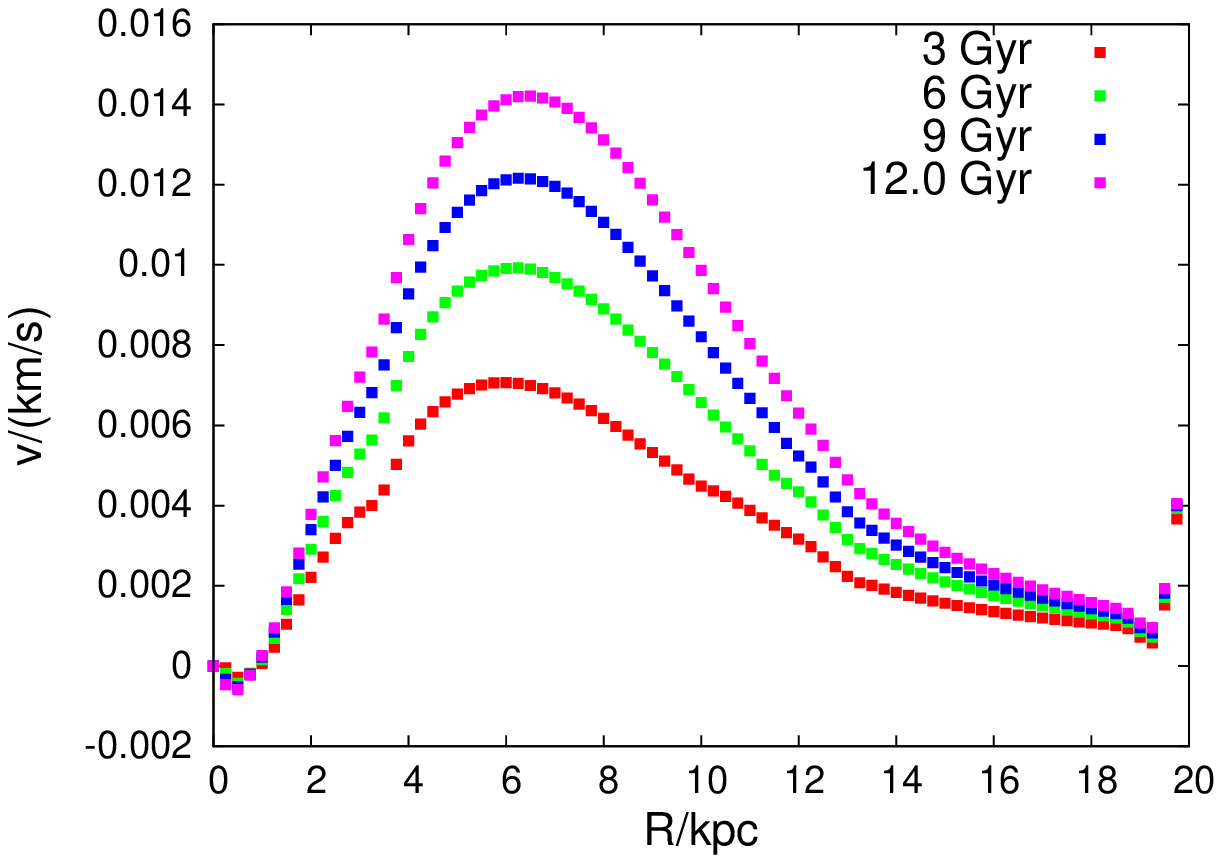}
 \caption{Flow velocities due to gas ejected from dying stars}
\label{fig:dyingstars}
\end{figure}

In Fig.~\ref{fig:dyingstars} the results are plotted in the form of a velocity map covering the complete simulation. In the intermediate disc regions the inflow (positive values) driven is strongest, since there is still significant stellar heat, the circular geometry is less important and so the effective asymmetric drift takes its maximum toll. The small outflow seen in the inner rings should not be taken too seriously as it is likely caused by a numerical instability (the relative angular momentum/radius change to reach the adjacent ring is far smaller for the outer neighbour, which is a mere effect of the grid size). This region is anyway uninteresting, since the bar is not modelled. The outer boundary produces a similar effect in the outermost rings where stronger inflows develop due to the fact that in the simulation no stars can come from beyond 20\eunit{kpc}.
The resulting flow velocities are of the order of $v \sim 0.01\eunitfrac{km}{s}$, negligible compared to the flow velocities onfalling material causes, which were of the order $v \sim 1.0\eunitfrac{km}{s}$. Therefore, their impact on the chemical evolution will not be significant and the conclusions of the last section remain unchanged apart from slight corrections, which are certainly smaller than the expected errors and uncertainties due to other processes. This statement may change if stars can exhibit a significant friction with the surrounding gas. We know that the Heliosphere extends far beyond the planetary system of our Sun and that many stars produce visible bow shocks in the interstellar medium. In case this interaction with the surrounding medium can slow down the relative motion of the stars and is not only an expression of the momentum balance with the stellar wind, the calculated terms could increase significantly.

\section{Inside-out Formation }\label{sec:inside-out}
In the following section we will implement a simple model of inside-out formation in the simulations. We will still assume an exponential gas density profile, but make the gas scale length a function of time. Then we have
\begin{equation}
    \Sigma_g(R)=\Sigma_{g,0}(t) e^{-R/R_g(t)} \, .                       
\end{equation}
Accordingly, in each timestep of the simulation a new density profile is calculated, from which the required amount of inflow and onfall follows by the flow prescriptions
\begin{align*}
 M_n &= 2\pi \Sigma_{0}(t)\int_{R_n}^{R_{n+1}} { e^{-R/R_g(t)} RdR} \\
     &= 2\pi \Sigma_{0}(t) R_g^{2}(t) \left[e^{-R/R_g(t)}\left( \frac{-R}{R_g}-1 \right) \right]_{R_n} ^{R_{n+1}}
\end{align*}
$\Sigma_{0}(t)$ is then determined by the total gas mass in the disc at this time. If $M_{gas}(t)$ is this total gas mass, then
\begin{equation*}
 \Sigma_{0}(t) = \frac{M_{gas}(t)}{\sum_n{M_n}}
\end{equation*}
where the sum runs over the annuli of the simulation.

The evolution of the scale length itself is neither tightly constrained by observations nor by theoretical predications. Again for simplicity we will assume it to be linear, i.e.
\begin{equation}
 R_g(t) = R_i + (R_f - R_i) \, t / t_f
\end{equation}
where $R_i$, $R_f$ and $t_f$ are the initial and final scale length and final time respectively. We will consider two cases, the first in which the gas scale length grows linearly over the whole duration of the simulations, the second in which the growth is limited to a time $t_f$ smaller than the present day disc age of 11.7\eunit{Gyrs} and constant thereafter. Note that in the second case the stellar scale length will still continue to grow after the point when the gas scale length remains constant because the stellar population needs time to adjust to the new scale.

In the reference model the gas scale length is constant in time at  $R_g = 3.5 \eunit{kpc} $ corresponding to a stellar disc scale length of $R_{star}= 2.5\eunit{kpc} = R_g / 1.4$ taking the exponent $1.4$ in the star formation rate into account. This value is in agreement with recent studies of the MW by \cite{Juric2008} who find $(2.6 \pm 0.52)$\eunit{kpc} for the thin disc. As constraints for the final and initial values we initialize the gas disc at $R_i = 2.80 \eunit{kpc}$ and let it grow up to $R_f = 3.70$\eunit{kpc} for $t_f = 11.7\eunit{Gyrs}$ in the first and $t_f = 4.5\eunit{Gyrs}$ in the second case. These values correspond to an initial stellar disc scale length of $R_{star} = 2.00 \eunit{kpc}$, consistent with the value of $(1.96 \pm 0.05)$\eunit{kpc} found by \cite{Bovy2011} for old stars in the MW.
The slightly higher final gas scale length is necessary when aiming to reach the stellar disc scale length of the reference model despite the lag of the stellar population behind the gas disc. Values significantly higher than the final scale length for the gas considered here seem to be unlikely, \cite{Kalberla2008} find a scale length of 3.75\eunit{kpc} for the MW. Moreover, we will see that the lag introduced by former star formation is so strong that a prolonged episode of significant gas disc growth seems hardly reconcilable with the currently accepted observations. The formation time for the second model was chosen such as to roughly reproduce the stellar scale length of the reference model at 11.7\eunit{Gyrs}. Then the stellar disc will have grown by 25$\%$ over the simulation time, as found for disc galaxies since $z=1$ \citep{Trujillo2005,Mateos2011}.

We consider these models to investigate the general effect of an inside-out formation on the simulations and our results for the angular velocity of the onfalling material. The treatment is certainly oversimplified and is not meant to model the details of the actual formation processes or give the correct time dependence. However, even with such a simple toy model we will at least see the most important trends.

As the inside-out formation changes the star formation history, to make the simulations better comparable to the reference model, we adjust the mass parameters $M_1$ and $M_2$ in the onfall law given by
\begin{equation*}
 \dot{M}=\frac{M_1}{b_1}\, e^{-t/b_1} + \frac{M_2}{b_2} e^{-t/b_2} \,
\end{equation*}
to yield similar results for the star formation at the solar radius. Recall that $b_1$ was a short time scale to model the build up of the Galaxy and $b_2$ a longer time scale responsible for sustained star formation. Specifically, the reference values are $M_1=0.45\times 10^{10}\, \msun$ and $M_2=2.9\times 10^{10}\, \msun$, the first inside-out simulation uses $M_1=0.75 \times 10^{10}\, \msun$, $M_2=2.85 \times 10^{10}\, \msun$ and the second $M_1=0.70\times 10^{10}\, \msun$, $M_2=2.75\times 10^{10}\, \msun$. The increased mass for the short timescale onfall serves to produce the early peak in the SFR and the mass for the long time scale is decreased to account for the additional gas mass due to a higher scale length at later times.

\begin{figure}
 \centering
 \includegraphics[width=0.45 \textwidth,angle=0,keepaspectratio=true]{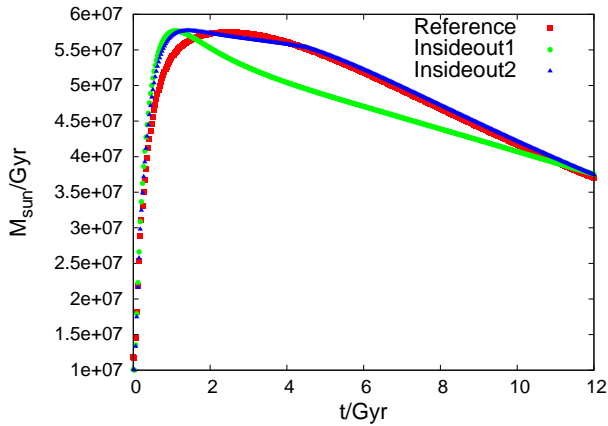}
 \caption{SFR at $R=7.5 \eunit{kpc}$ over the simulation time for the reference model and the two inside-out scenarios }
 \label{fig:inside-out_sfr}
\end{figure}

The corresponding star formation rates as functions of time at $R=7.5 \eunit{kpc}$ are shown in Fig~\ref{fig:inside-out_sfr}. As explained the parameters are chosen such that they roughly show a comparable evolution. In the case of the first inside-out simulation this is not entirely possible due to the ongoing growth of the gas disc over the whole time. The lower gas scale length redistributes the mass towards the centre of the Galaxy and to keep the SFR high enough farther out would require considerably higher total masses in the models. We decided to keep the final SFR comparable which leads to a deficit at earlier times as the scale length of the gas disc is smaller. For the second inside-out simulation the agreement is qualitatively better. In both cases the final and peak values of the SFR agree fairly well, but the peak is shifted towards earlier times. Without changing the functional dependence of the infall law itself a better fit can hardly be achieved.

\begin{figure}
 \centering
 \includegraphics[width=0.45 \textwidth,angle=0,keepaspectratio=true]{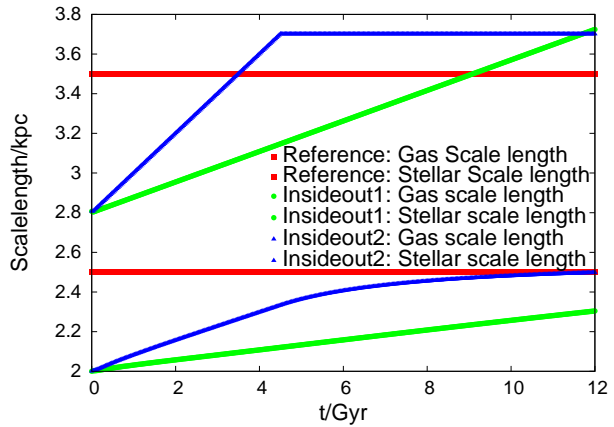}
 \caption{Gas and stellar disc scale lengths over the simulation time for the reference model and the two inside-out scenarios}
 \label{fig:inside-out_scalelength}
\end{figure}

Next we consider how the gas scale length affects the stellar scale length of the simulations shown in Fig~\ref{fig:inside-out_scalelength}. The first model shows a linear increase of the stellar scale length from the initial value of 2.0\eunit{kpc} to a final value of 2.30\eunit{kpc} failing to reach the reference value. Continuing the simulations for another 1.8\eunit{Gyrs} giving a gas and stellar scale length of 2.34\eunit{kpc} and 3.84\eunit{kpc} respectively is still not sufficient to remedy this problem. Looking at the second model we see a growth of the stellar scale length from the initial 2.0\eunit{kpc} to the final 2.5\eunit{kpc}. Here the growth is linear up to a short time after the gas disc stops growing at 4.5\eunit{Gyrs}, and then slows down but does not stop. In both simulations stars born at earlier times and smaller scale lengths survive to the present time and consequently reduce the scale length of the total stellar population. As the SFR peaks during the first 3\eunit{Gyrs}
stars born at that time contribute heavily at all later times.

These problems can certainly be solved by assuming a different functional form for the time dependence of the gas scale length. However, this can only be accomplished by increasing the scale length at early times to reduce the contribution of stars born at low scale lenghts. The second model serves to illustrate how it then becomes possible to reach the reference value for the scale length at the final time. To achieve the same in the first model would require unrealistically high values for the final gas scale length of 4.3\eunit{kpc} in conflict with observations.

\begin{figure}
 \centering
 \includegraphics[width=0.45 \textwidth,angle=0,keepaspectratio=true]{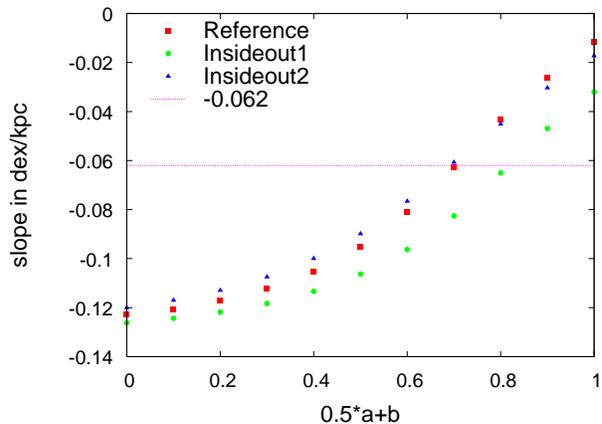}
 \caption{Metallicity gradient of the simulations as a function of the average angular velocity of the onfalling gas for the reference model and the two inside-out formation scenarios}
 \label{fig:inside-out_slope}
\end{figure}

Finally we turn to consider the effect of the inside-out formation on the metallicity gradient. Generally, an inside-out formation can steepen the gradient. This expectation is only partly true for our simulations as seen in Fig~\ref{fig:inside-out_slope}. For the first simulation the gradient is indeed steeper for all the parameter values considered. As the gas has a shorter scale length over long parts of the simulation compared to the reference simulation, more stars and consequently metals are produced in the inner than in the outer parts of the Galaxy adding to the gradient produced by the flows. For the second simulation two competing effects influence the metallicity gradient. With little to no flows ($0.5 a +b > 0.6$) the inside-out simulation has a steeper gradient, in this case the initial formation period and the gradient created by a shorter scale length of the gas dominates over the effects of the flows. But in the parameter regime of high flows, the inside-out formation actually reduces the
flows as the inner regions of the galaxy already have sufficient mass due to a shorter scale length and do not need high radial flows to sustain their mass profile.

In summary, an inside-out formation scenario has notable effects on the metallicity gradient, the first case would approximately yield $\Delta v/\Vc = 0.1$. But the interplay between radial flows and the formation process complicate matters in such a way that the gradient does not always steepen as in the second simulations. In particular, we note that a simple linear growth of the gas scale length amplifies the effects of the inside-out formation and does not seem plausible considering the high final gas scale length necessary. An inside-out formation that mainly occurs at early times and slows down has considerably smaller effects on the metallicity gradient.

\section{Origin of Infall}\label{sec:origin}

So far we have discussed the implications of the coupling of radial inflows to an assumed accretion onto the Galactic disc and explored the behaviour of the model under different assumptions on the parameters and underlying laws. But the question of the origin of the accreted material remained open. Different theories exist and our simulations show values for the kinematic properties of the onfalling material comparable to those predicted by an interaction of gas clouds with a rotating corona. However, it is widely believed that extended gaseous outer disc like that of the Milky Way \citep{Kalberla2008}, are the dominant reservoir and source to nourish the inner disc's star formation \citep[see e.g.][]{Sancisi2008}. So far it is not clear how this gas should loose its angular momentum to move inwards, but assuming such a mechanism was found, the consequence would be rather high radial flows of gas through the edge of the stellar disc. The implications of these radial flows on the chemical evolution of the
Galaxy have not been considered before, but might be used to put limits on the possible accretion rates.

As we saw before, by the intimate relationship between radial flows and radial abundance gradients shown e.g. in Fig.~\ref{fig:completegrid_slope} the observed abundance gradient imposes strong limits on the maximum flows and hence on the maximum ``through-disc'' accretion of our Galaxy.

To estimate the effect of inflow from such extended gas layers we compare the total accreted masses during the past $3 \eunit{Gyrs}$, i.e. in the time from  8.7\eunit{Gyrs} up to 11.7\eunit{Gyrs} by onfall onto the disc and by inflow into the outermost ring. We take the total mass accretion in the outermost ring as a good measure for the through-disc inflow.

\begin{figure}
 \centering
 \includegraphics[width=0.49 \textwidth,angle=0,keepaspectratio=true]{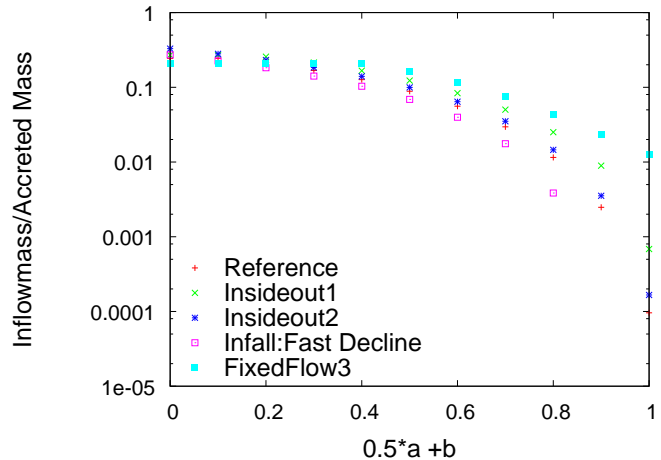}
  \caption{Upper panel: the fraction of total inflow mass to accreted mass on a logarithmic scale as a function of average angualar velocity $\bar{v}$, for the reference model, the two inside-out formation scenarios, the model with a fast decline of infall and the model with an additional fixed inflow.}
  \label{fig:accretion_comp}
\end{figure}

The sum of accreted mass and inflowing mass is independent of the flow parameters as in our implementation flows do not change the mass profile of the disc. But as higher flows build up throughout the disc the relative importance of radial inflow into the edge of the disc should increase. In Fig.~\ref{fig:accretion_comp} the mass fraction of inflowing mass to accreted mass for the last 4\eunit{Gyr} is shown for the reference simulation, the two inside-out scenarios from the preceding section, a model with a steeply declining infall rate, and the model with an additional fixed flow velocity of 0.6\eunitfrac{km}{s}. Recall that the first inside-out scenario had a continuously growing gas disc scale length, while the second only has a growth period in the first 4.5\eunit{Gyrs}, and both models end up with a 
higher gas scale length than the reference model. The infall model has a faster exponential decay of the mass of infalling material, specifically a decay time of 7\eunit{Gyrs} instead of 14\eunit{Gyrs}. For all these models the mass fraction drops as $\bar{v}$ increases and flow levels across the whole disc and at the edge decrease. The inside-out simulations generally show a higher mass fraction than the reference simulation. As the gas scale length at the end of the simulation was in both cases higher than in the reference model, the mass in the outermost ring is higher as well and more mass
flows in at the edge of the Galaxy. Additionally, the inside-out growth of the disc causes higher flows in the outer regions of the Galaxy and, thus, a higher flow into the last ring. The second inside-out scenario is again closer to the reference value as most of the growth occurred before the time-frame over which we add the masses, and only the effect of the higher gas scale length is relevant in this case. The model with a faster decline of the infall rate shows a smaller mass fraction as total flow levels are reduced compared to the reference simulation. Finally the model with an additional constant flow velocity shows a considerably higher mass fraction for most parameter values. This is easily understood as this model generally has higher flow levels and always drives a radial flow in the outermost ring, which is not true for the models without this additional flow. Again the lowest values of $\bar{v}$ should not be considered for this model.

To compare the simulations, note that the model with additional fixed inflow should be offset by roughly $\Delta v / \Vc = +0.15$ and the model with a fast decline of the infall by $\Delta v / \Vc = -0.05$ to yield similar metallicity gradients. At $\bar{v} = 0.5$ the first inside-out scenario yields a contribution of inflow of 11 \%
to the total accreted mass, for the other simulations the contribution at comparable values of $\bar{v}$ is even lower. This value corresponds to a metallicity gradient shallower than $-0.095 \eunit{dex/kpc}$ for the reference model. Even considering some of the possible uncertainties discussed above $\bar{v} >= 0.5 $ seems to be a reasonable lower bound on the average angular velocity of onfalling material.

We conclude from these results that radial accretion from the outer disc gas reservoir through the Galactic disc should be relatively small. Considering a moderate lower bound of $\bar{v} >= 0.5 $, the radial accretion contributes less than 11 \% to the total accreted mass in the models considered here. A comparable participation (fraction $> 10\%$) of the outer gaseous disc in feeding the disc would create radial abundance gradients far in excess of the observations. A mechanism for the material to lose most of its angular momentum and to distribute the material over the whole disc without causing flows within the disc, i.e. flowing above the disc and without participating in the advection of metals towards the centre would be required to be in agreement with the observed metallicity gradient. Yet, currently such a construction appears artificial and lacks any observational evidence.

\section{Conclusions}\label{sec:sum}

While recent works concentrated mostly on describing the effects of inflows of a certain preset velocity, we coupled the inflow process to conservation of angular momentum between the accreted material and the gas already present in the disc. In this picture, the angular momentum of the disc gas is diluted by freshly accreted material with lower specific angular momentum driving a radial flow through the disc. This coupling is used to derive the spatial structure of the radial flows consistently within the model. The resulting flow and onfall pattern is not known from the beginning, but rather a prediction of the model. Due to its physical motivation the new prescription is preferable to the older prescription of SB09 and also to models with a preset inflow.

We have shown that the chosen prescription can produce a vast range of radial metallicity distributions in disc galaxies and studied the dependence of the observables on different assumptions about the qualities of the inflow.

We fitted the model to observations of the present day metallicity gradient of our Galaxy by \citet{Luck2011II}. This comparison constrained the average rotational velocity of the onfalling material to be $ 0.7 \le v/\Vc \le 0.75$ corresponding to $154\eunitfrac{km}{s} \le v_{onfall} \le 165 \eunitfrac{km}{s}$ assuming a constant $\Vc = 220 \eunitfrac{km}{s}$.
This value compares well with the predictions for an accretion of gas from the corona (when neglecting vertical velocity gradients) at about $V_{corona} \approx 145\eunitfrac{km}{s}$ \citep{Marinacci2011, Marasco2011}. However, the specific accretion profile predicted by \cite{Marasco2011} is not compatible with our simulations, since the rather sharp peak in Galactocentric radius would drive flows that create a radial metallicity distribution at odds with the data, if not some other process fills the gap especially in the inner disc regions.

One major uncertainty is how much angular momentum is given up by the gas itself through various channels. We studied a so far neglected process where the ejection of material from stars moving on non-circular orbits dilutes the angular momentum via their asymmetric drift. If there is not an unknown amplification, this momentum exchange has only a very small impact on radial flow velocities of the order of $v_{flow} \sim 0.01\eunitfrac{km}{s}$ and can be safely neglected. More of a concern are the well-known processes of gravitational interaction of the gas with spiral arms or the bar and possibly friction within the gas layers. Considering the estimates by \cite{LaceyFall1985} they might play a role in additionally changing the gradient. We have considered models in which a fixed constant flow velocity was added to our flow model to describe part of the uncertainty of these processes. We conclude that flows at that level certainly change our results quantitatively if they extend across the whole Disc and
persist for the whole age of the Galaxy, but for values within the estimates the total effects remain small of order $0.05 \, v/\Vc$. However, the local structure and gas dynamics may well be affected considerably without affecting the global flow levels and the gradient.

Resonances stable in galactocentric radius are another source of uncertainty to the flow pattern. Any significant influence, however, should show up as an excursion in the radial abundance profile and up to now is unobserved, making such a bias a minor concern.

Radial migration \citep{SellwoodB01} has only a minor influence in flattening the inner disc abundance gradient slightly via mixing. Even larger amplitudes are easily compensated by the inflow.

These results were shown not to depend strongly on the resolution or the churning amplitude. Additional dependence on model parameters, specifically the assumed infall history and the star formation efficiency, was found to be relatively small.

Declining star formation rates tend to reduce the gradient in our models. As the radial flows are directly coupled to the accreted material which determines the star formation rates, a declining SFR also means decreasing flows. For the specific model parameters, a very fast decline of the SFR shifts the gradient to lower values, but a slower decline has relatively little effect in steepening the gradient.

The star formation efficiency mostly affects the simulation by controlling the total amount of gas in the disc. For different levels of radial flows, the effects on the gradient are qualitatively different, but for the region corresponding to the observed data a higher gas mass tends to flatten the gradient, hence needing stronger flows and a lower angular momentum of onfalling material to give the correct value.

We also examined the impact of inside-out formation. Generally inside-out formation raises the ratio of metal production to infall of fresh material in the inner disc relative to the outer disc facilitating a steeper metallicity gradient. However, a simple linear growth of the gas scale length over the complete age of the Galaxy is strongly limited by observations, and a model in which the gas disc mainly grows in the early times of the simulations is more adequate in order to reproduce the stellar disc scale length measured today. In this latter case the effects on the present-day metallicity gradient are greatly reduced. It was impossible for us to achieve the measured abundance gradient just by inside-out formation and so conclude that radial flows are essential for explaining the observations.

Independent from these considerations we showed that the large amounts of HI gas in extended layers beyond the stellar disc cannot feed the majority of fresh material required to the inner disc, since this would imply too large radial flows and a too steep abundance gradient.

\section*{Acknowledgements}
We thank the anonymous referee for very helpful comments. T.B. acknowledges financial and material support from Max Weber-Programm, R.S. acknowledges support from Max-Planck-Gesellschaft and by NASA through Hubble Fellowship grant $\#60031637$ awarded by the Space Telescope Science Institute, which is operated by the Association of Universities for Research in Astronomy, Inc., for NASA, under contract NAS 5-26555.

\section{Appendix: Resolution Effects}\label{subsec:reseff}

\begin{figure*}
\centering
 \includegraphics[width=0.32 \textwidth,angle=0,keepaspectratio=true]{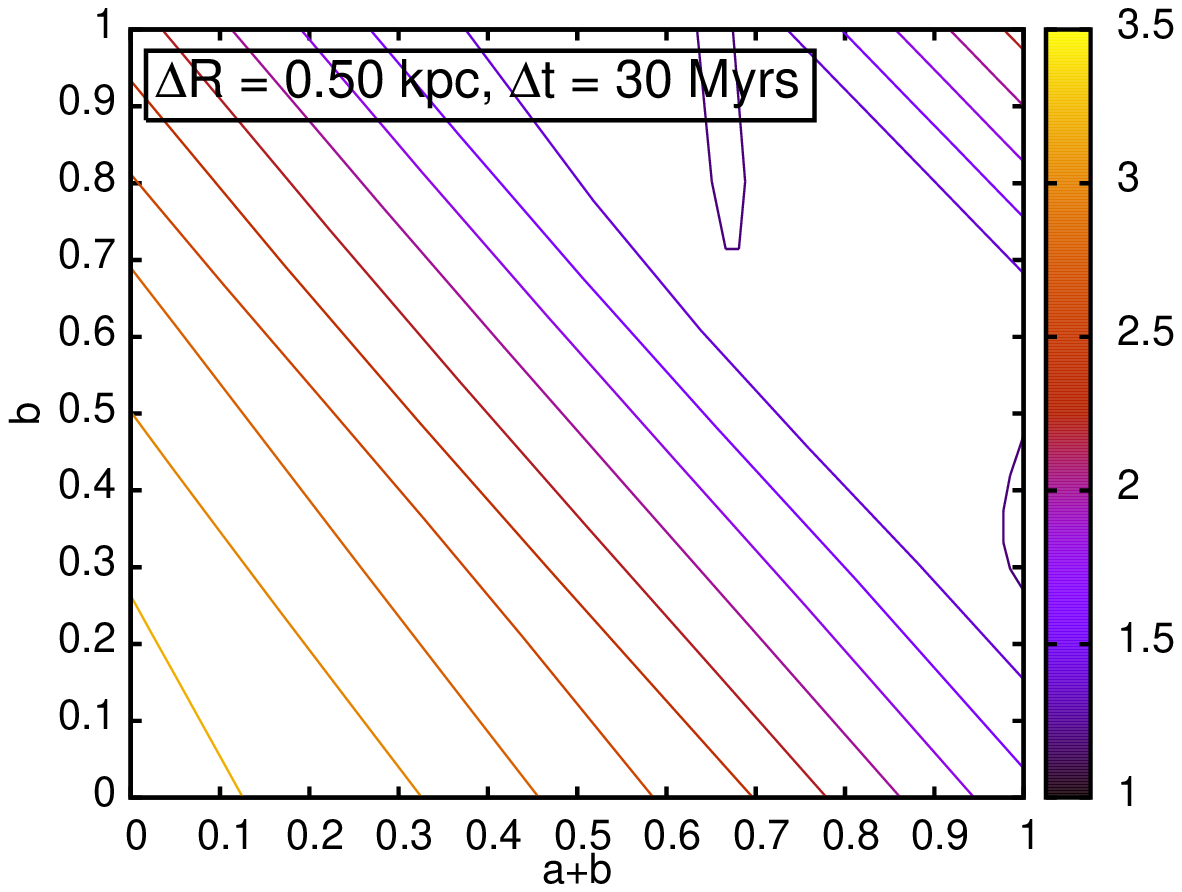}
 \includegraphics[width=0.32 \textwidth,angle=0,keepaspectratio=true]{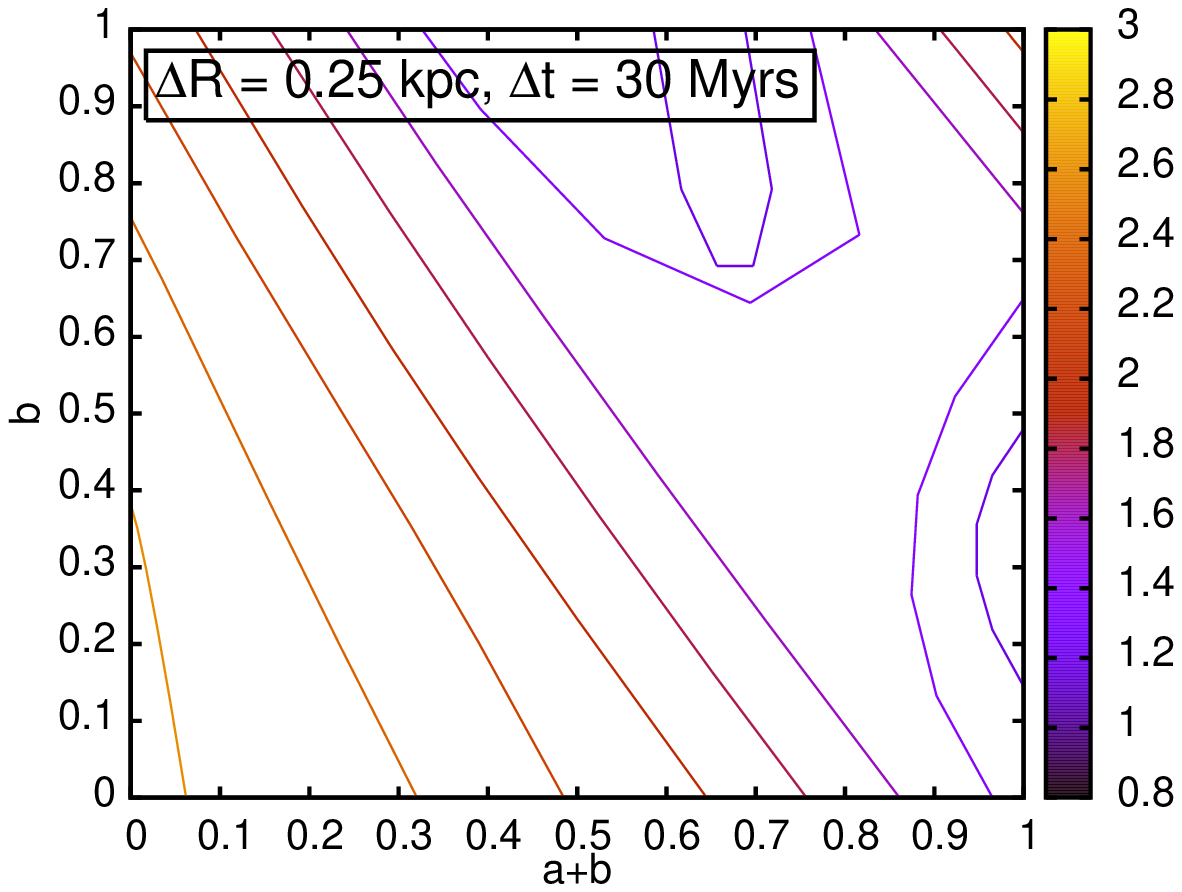}
 \includegraphics[width=0.32 \textwidth,angle=0,keepaspectratio=true]{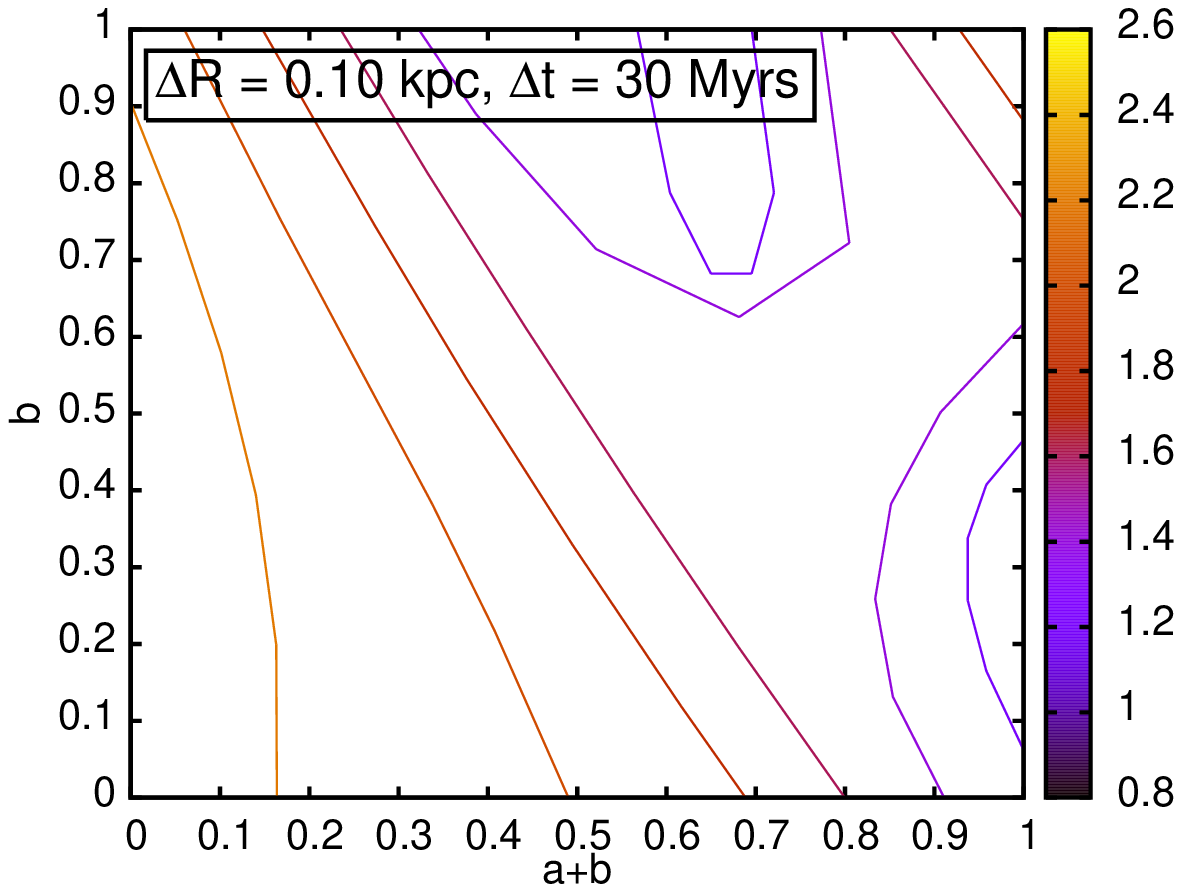}
\caption{${\chi}_{red}^{2}$-contours for comparison with Luck11II Data on the coarse simulation grid for low (left), standard (middle) and high (right) space resolution}
\label{fig:res_chi2_comparison}
\end{figure*}

\begin{figure*}
\centering
 \includegraphics[width=0.32 \textwidth,angle=0,keepaspectratio=true]{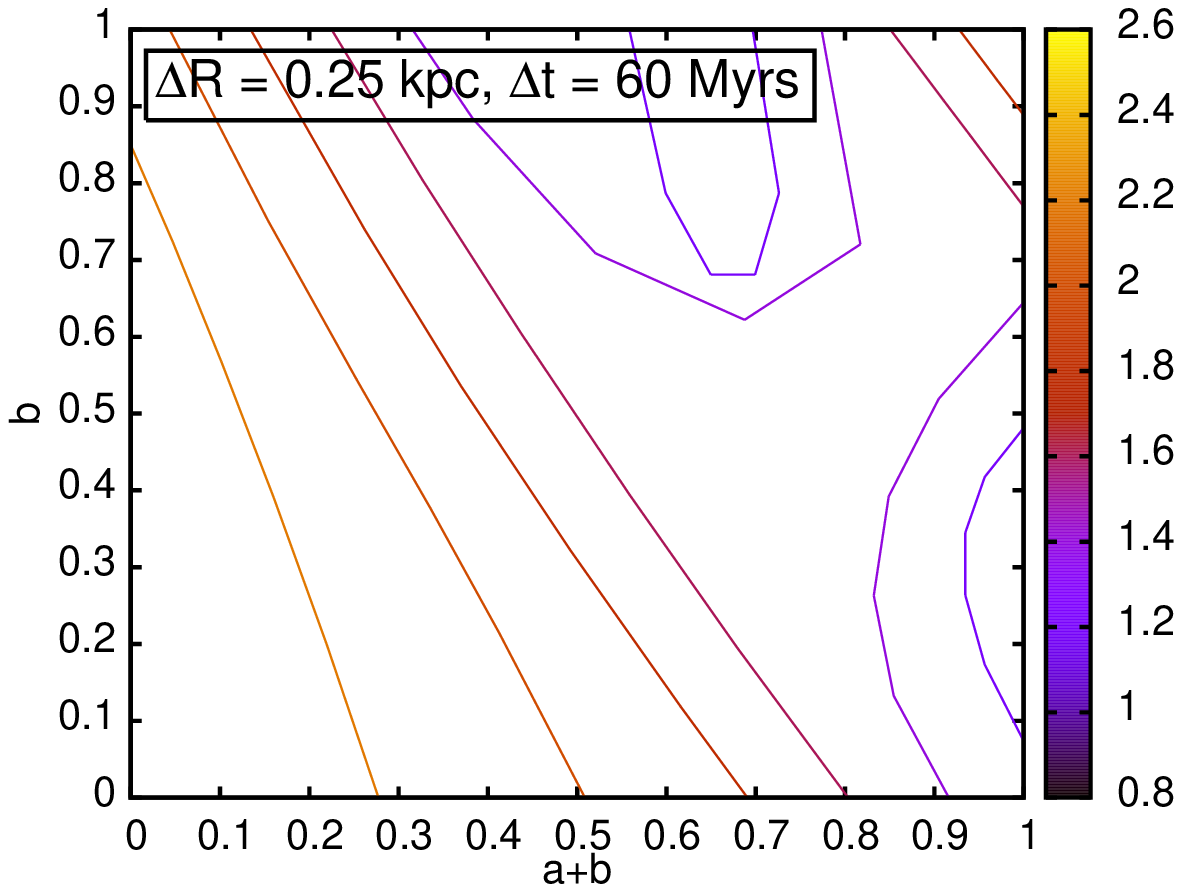}
 \includegraphics[width=0.32 \textwidth,angle=0,keepaspectratio=true]{./graphs/11_II_Reference_chiprobcontours.eps}
 \includegraphics[width=0.32 \textwidth,angle=0,keepaspectratio=true]{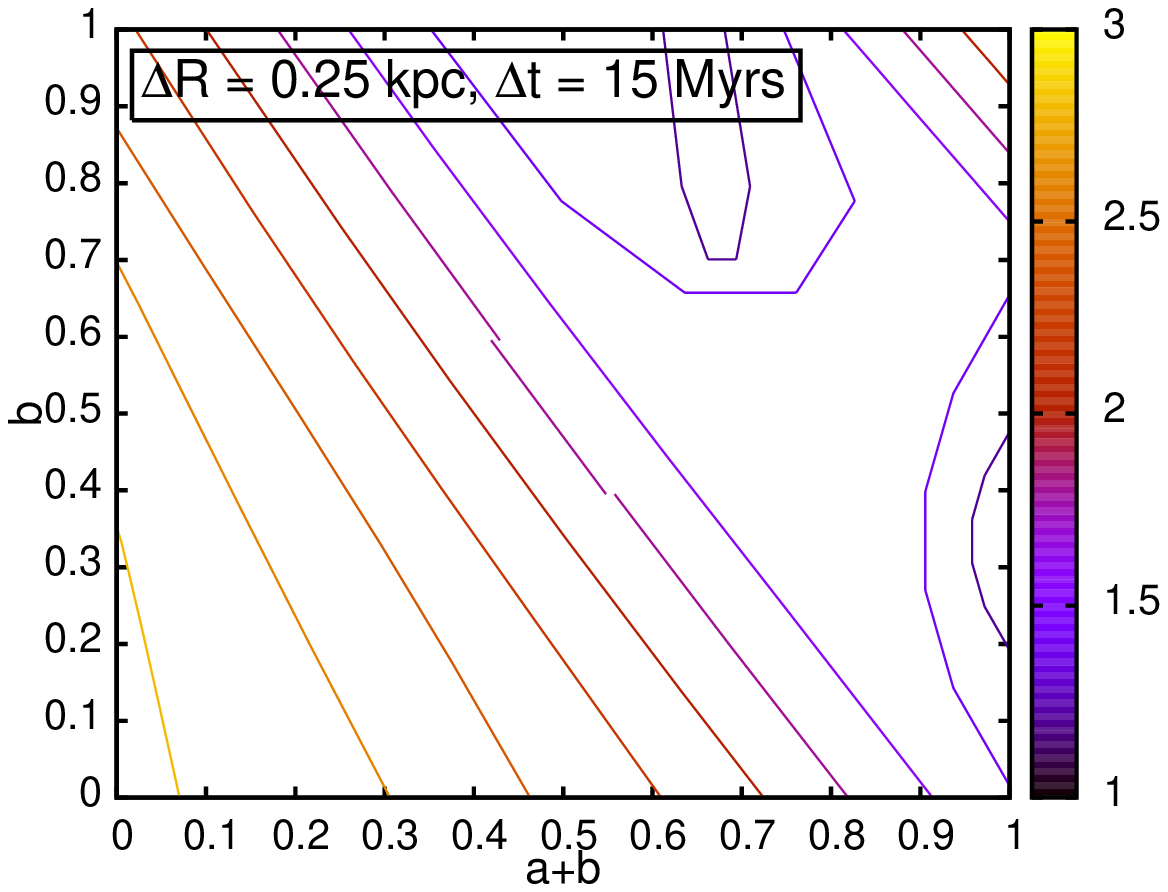}
\caption{${\chi}_{red}^{2}$-contours for comparison with Luck11II Data on the coarse simulation grid for low (left), standard (middle) and high (right) time resolution}
\label{fig:timeres_chi2_comparison}
\end{figure*}

The resolution of the simulation has several impacts on the results, for example by ``numerical mixing'' on the flow, when the flow does not cross the width of one annulus in the simulation per timestep and hence the gas thought to be on a ring boundary mixes with the entire ring. For our tests we vary both the spatial and the time resolution separately by computing in addition to the standard parameters $\Delta R = 0.25 \kpc$, $\Delta t = 30 \Myr$ a coarse grid of models with $\Delta R = 0.1 \kpc$ and $\Delta R = 0.5 \kpc$ as well as $\Delta t = 15 \Myr$ and $\Delta t = 60 \Myr$. Further resolution effects result from the maximal flow velocity limit given by the grid size $v_{flow}<\Delta R/\Delta t$, which has some limited influence in the earliest timesteps.
For the resolutions used this corresponds to $v_{flow} < 3.3\eunitfrac{km}{s}, \, 8.2\eunitfrac{km}{s},\,16.4\eunitfrac{km}{s}$ for the higher space, normal and lower space resolution respectively and $v_{flow} < 4.1\eunitfrac{km}{s}, \, 8.2\eunitfrac{km}{s},\,16.4\eunitfrac{km}{s}$ for the high, normal and low time resolution simulations. Tightening the velocity limit can artificially improve the fits in the low infall angular momentum region, where the flows are too fast, but in the interesting region has virtually no impact. We demonstrate this in \figref{fig:res_chi2_comparison} and \figref{fig:timeres_chi2_comparison}. While both coarse resolution simulations show some minor deviation from the general picture, using higher resolution apparently brings no further benefits.

\begin{figure*}
 \centering
  \includegraphics[width=0.325 \textwidth,angle=0,keepaspectratio=true]{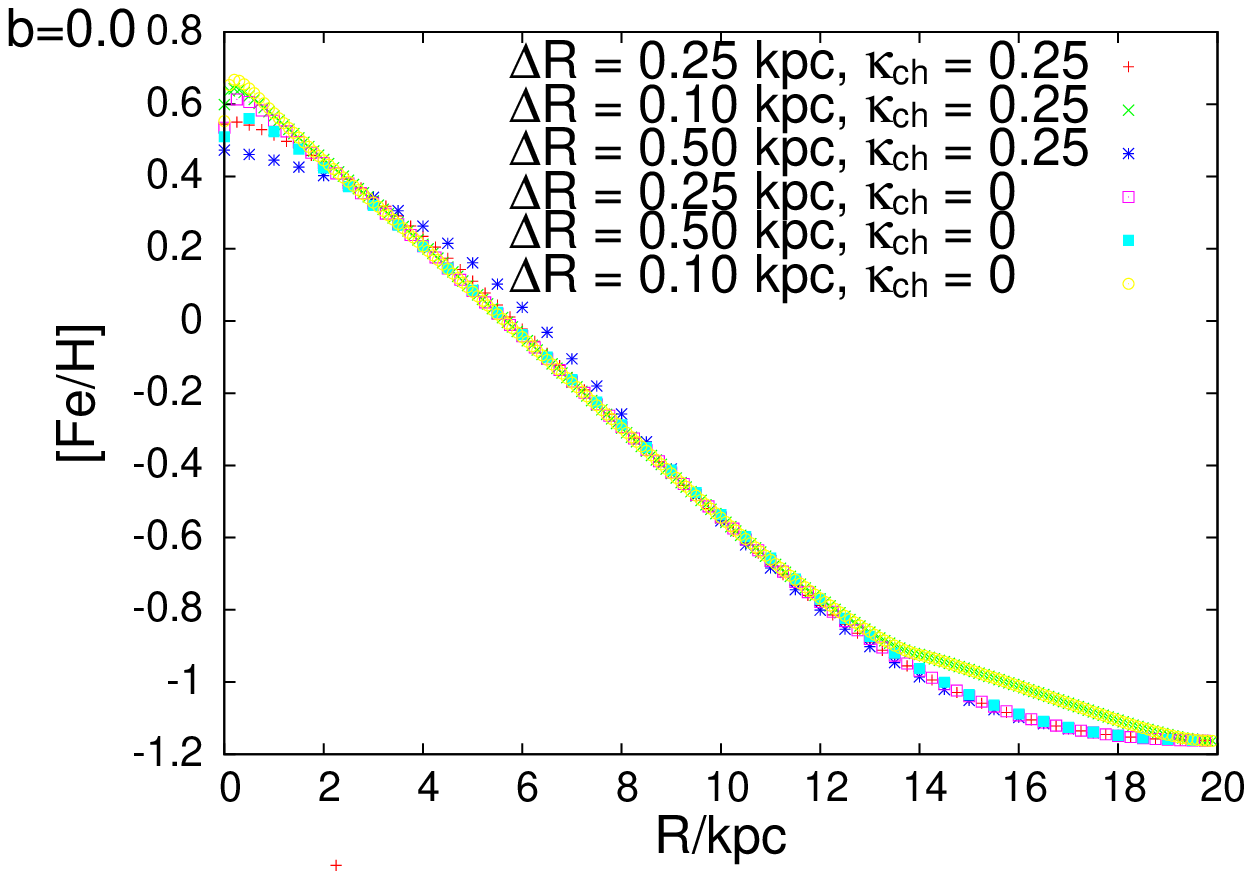}
  \includegraphics[width=0.325 \textwidth,angle=0,keepaspectratio=true]{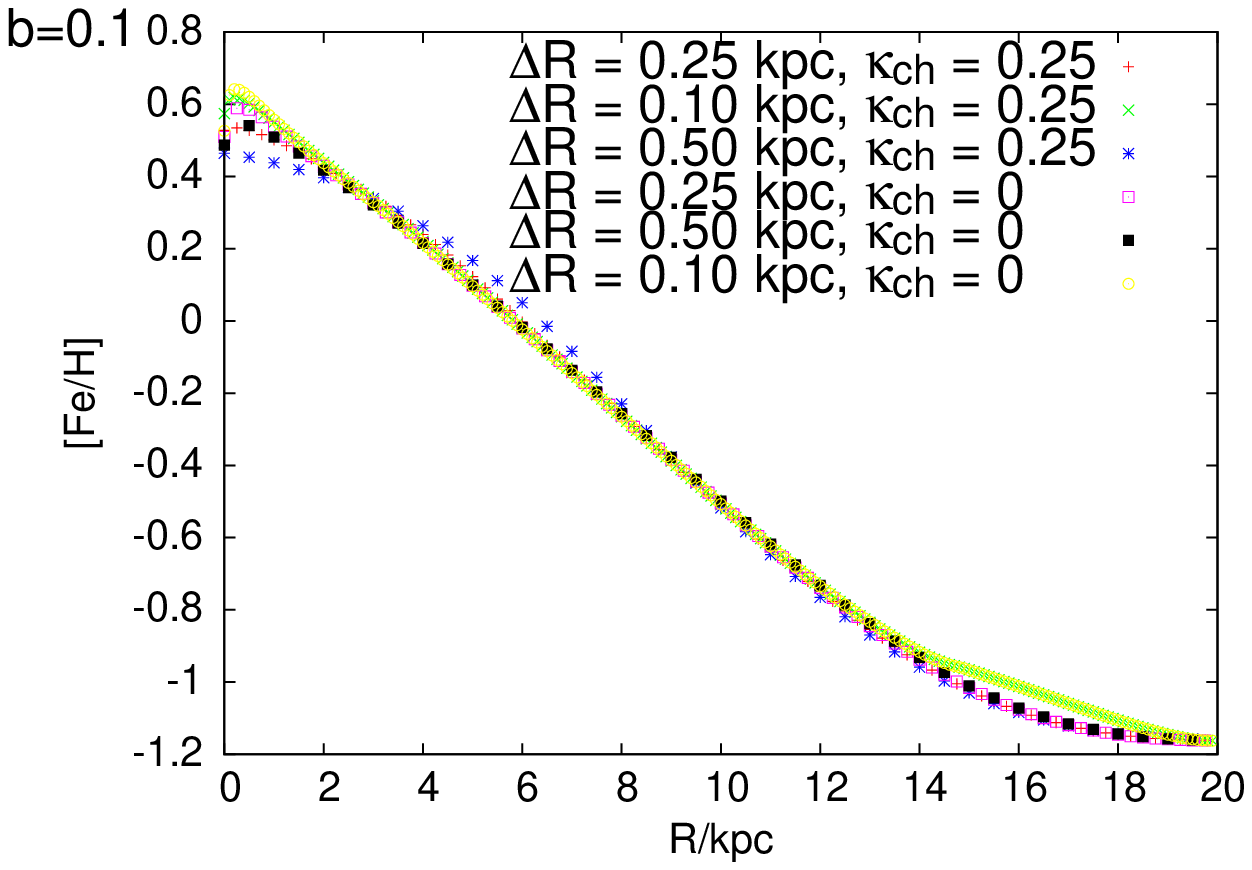}
  \includegraphics[width=0.325 \textwidth,angle=0,keepaspectratio=true]{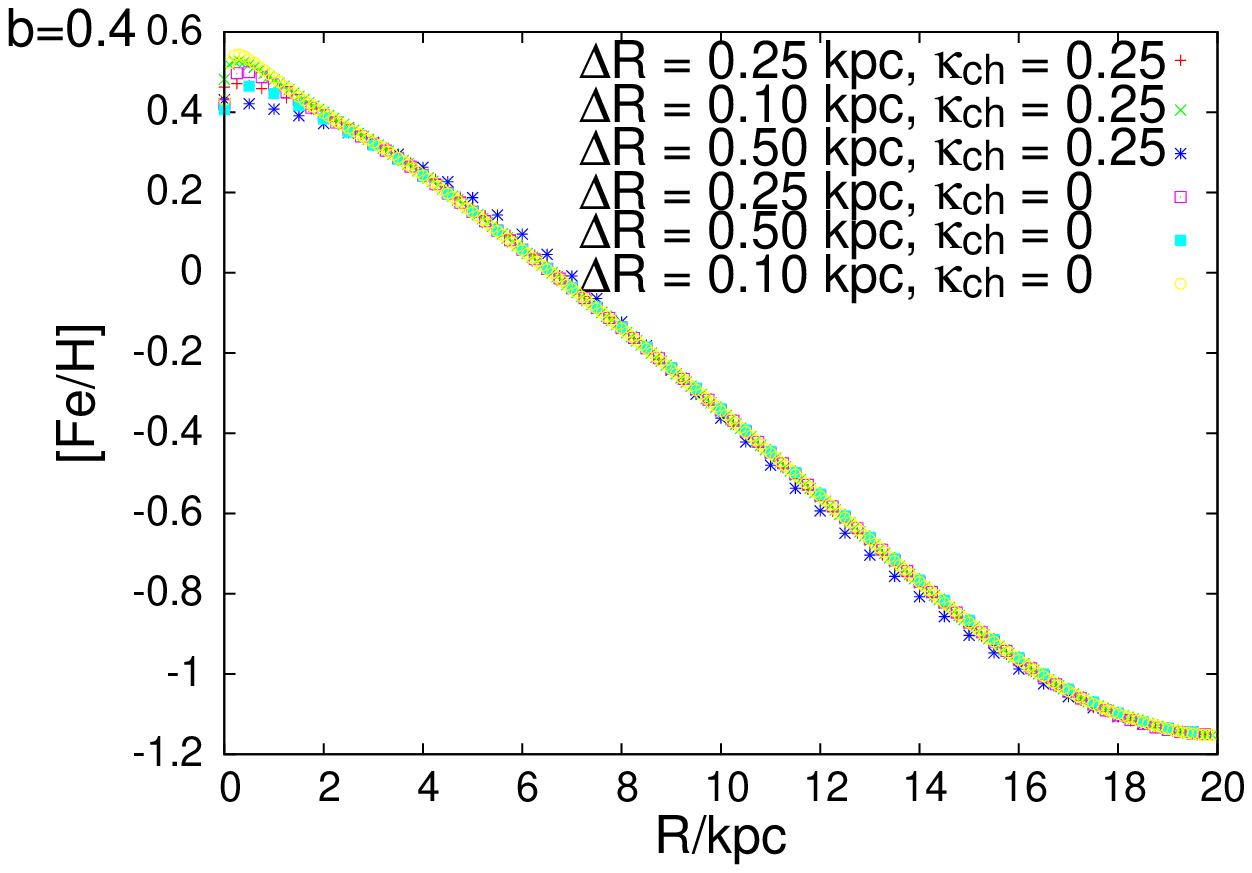}
  \includegraphics[width=0.325 \textwidth,angle=0,keepaspectratio=true]{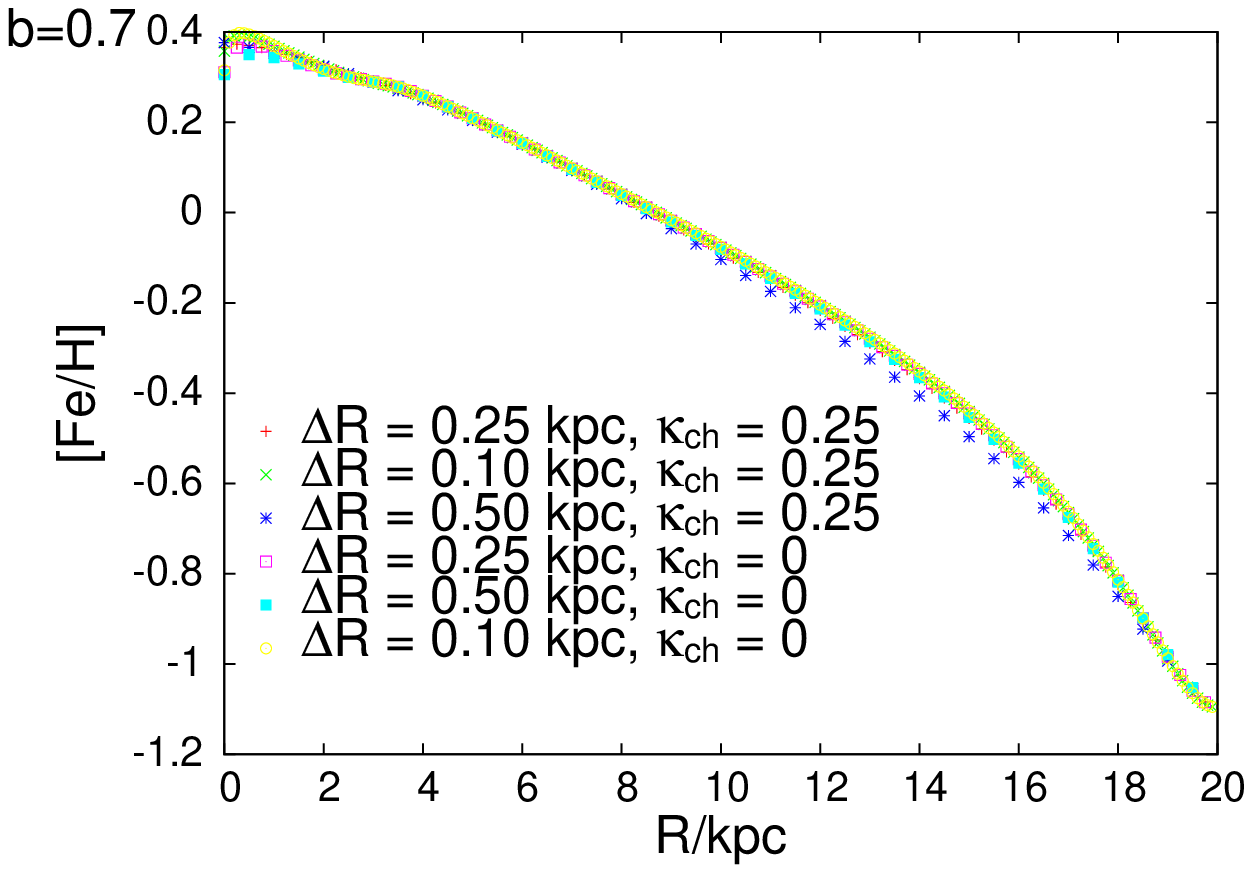}
  \includegraphics[width=0.325\textwidth,angle=0,keepaspectratio=true]{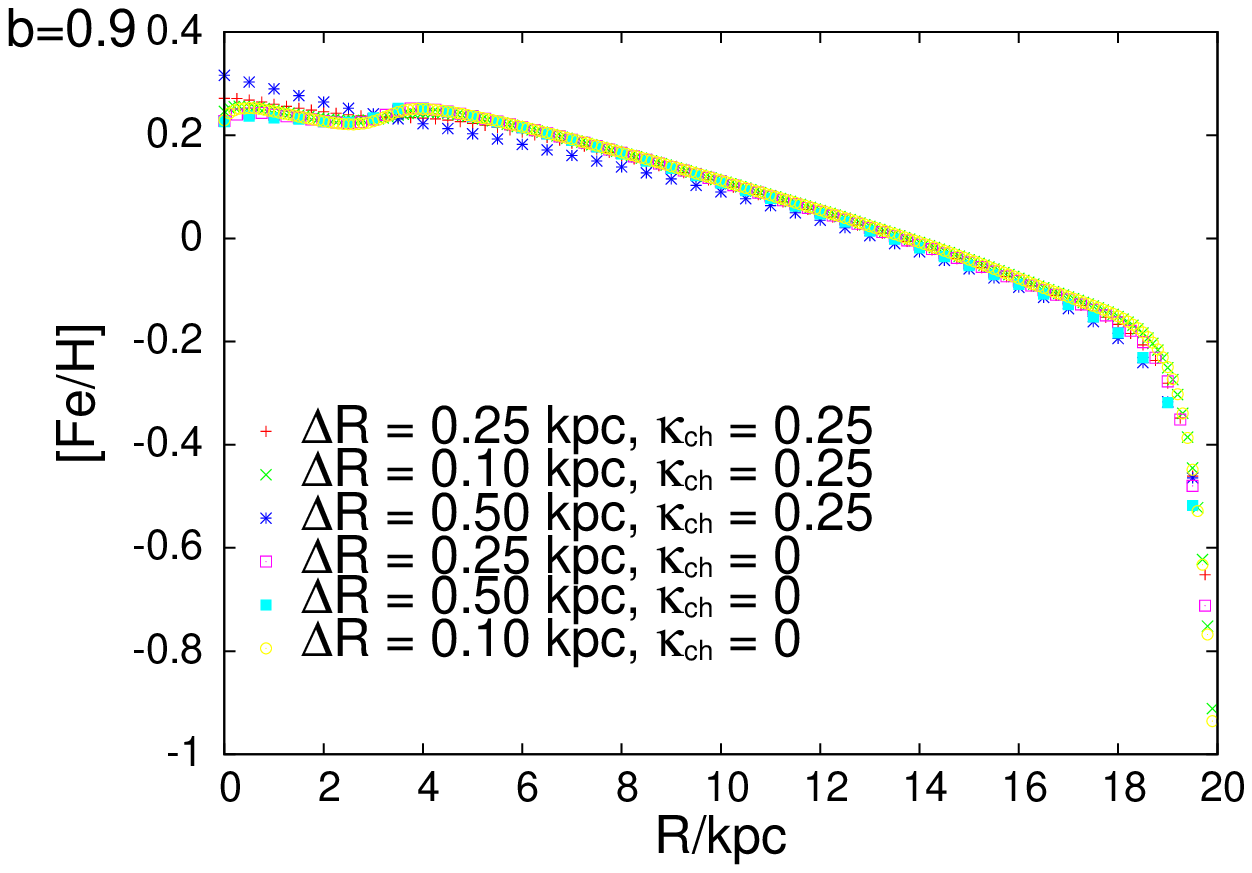}
  \includegraphics[width=0.325 \textwidth,angle=0,keepaspectratio=true]{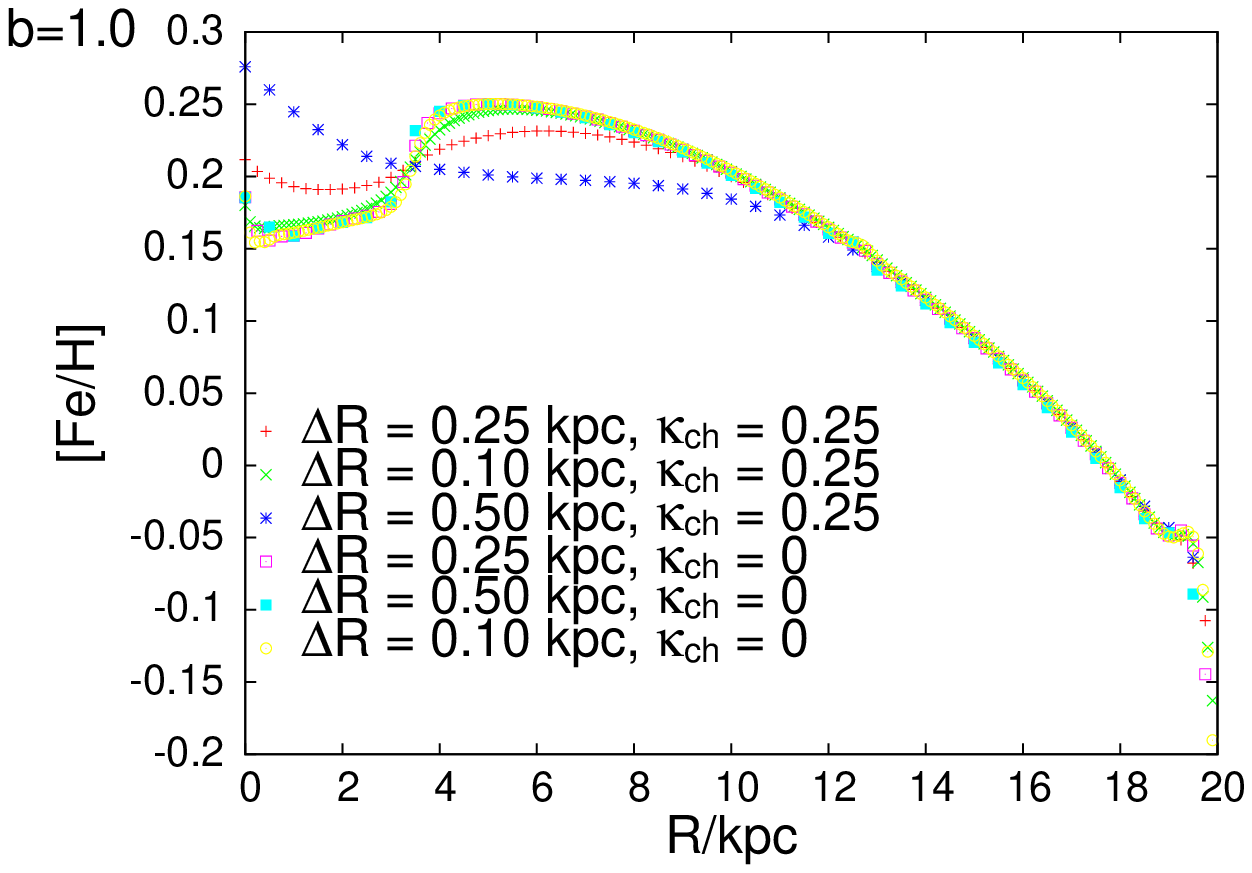}
 \caption{Resulting abundance gradients for models with different space resolutions and with or without churning, the flow parameters are $a=0$ and $b$ is given at the upper left corner of the plots}
 \label{fig:res_metallicity_comparison}
\end{figure*}

To test the effects of the churning on the results simulations with $a=0$, $b=0,0.1,\cdots,0.9,1$ were computed at identical resolution with and without churning. In its current implementation churning also involves the cold gas phase and directly mixes the gas apart from the stars and hence their yields being scattered over different radii. The results are shown in Fig.~\ref{fig:res_metallicity_comparison}. The only significant difference happens in the case ($b = 1.0$), where there are no systematic radial gas flows, which otherwise dominate any effect from churning. In other cases, churning mixes especially the inner disc and by this mildly lowers the inner disc radial abundance gradient. If churning was hence significantly stronger than set in \cite{SBI}, we need a bit lower inflow velocities or vice versa a stronger radial flow to obtain the same gradient. The effect is close to being negligible, however. We also note that in the formulation we used, churning gets stronger for coarse spatial resolution
and fine time resolution.

\label{lastpage}
\end{document}